\documentclass[manuscript]{aastex}

\shorttitle{Collisions between  sintered  icy aggregates}
\shortauthors{Sirono}

\begin{document}


\title{Collisions between  sintered  icy aggregates}


\author{Sin-iti Sirono}
\affil{Earth and Environmental Sciences, Nagoya University, Tikusa-ku, Furo-cho, Nagoya 464-8601 Japan}

\author{Haruta Ueno}
\affil{Earth and Environmental Sciences, Nagoya University, Tikusa-ku, Furo-cho, Nagoya 464-8601 Japan}

%\author{C. D. Biemesderfer\altaffilmark{4,5}}
%\affil{National Optical Astronomy Observatories, Tucson, AZ 85719}
%\email{aastex-help@aas.org}

%\and

%\author{R. J. Hanisch\altaffilmark{5}}
%\affil{Space Telescope Science Institute, Baltimore, MD 21218}


%\altaffiltext{1}{Visiting Astronomer, Cerro Tololo Inter-American Observatory.
%CTIO is operated by AURA, Inc.\ under contract to the National Science
%\altaffiltext{2}{Society of Fellows, Harvard University.}
%\altaffiltext{3}{present address: Center for Astrophysics,
%    60 Garden Street, Cambridge, MA 02138}
%\altaffiltext{4}{Visiting Programmer, Space Telescope Science Institute}
%\altaffiltext{5}{Patron, Alonso's Bar and Grill}


\begin{abstract}
Collisions between sintered icy dust aggregates are numerically simulated. If the temperature of an icy aggregate is sufficiently high, sintering promotes molecular transport and a neck between adjacent grains grows. This growth changes the mechanical responses of the neck. We included this effect to a simulation code, and conducted collisional simulations. For porous aggregates, the critical velocity for growth, below which the mass of an aggregate increases, decreased from 50\,m\,s$^{-1}$ for the non-sintered case to 20\,m\,s$^{-1}$. For compacted aggregates, the main collisional outcome is bouncing. These results come from the fact that the strength of the neck is increased by sintering. The numerical results  suggest that the collisional growth of icy grain aggregates is strongly affected by sintering. 
\end{abstract}





\section{Introduction}

Planetary formation starts from the collisional coagulation of sub-micron-sized dust grains which in turn leads to the formation of dust aggregates. These aggregates grow further through collisional sticking. Although the formation process of planetesimals is still unclear, collisional growth of dust grains and dust aggregates plays a critical role in the first step of planetary formation. 

Collisions of dust grains have been studied both experimentally and theoretically. An important parameter is the critical velocity for sticking, below which a grain (or grain aggregate) can stick to another. \cite{Chokshi} derived a critical velocity for sticking in collisions between two spherical dust grains. If the collisional velocity is higher than the critical velocity, the grain bounces. They determined that the critical velocity for silicate dust grains 1\,$\mu$m in size is $\sim$0.1\,m/s. On the other hand, the critical velocity determined experimentally is $\sim$100\,m/s \citep{Poppe}. Plastic deformation\citep{Poppe}, {\bf hysteresis of adhesion \citep{Krijt13}}  or a large surface energy \citep{Kimura} are possible causes of the discrepancy. Collisional growth is highly dependent on the material properties. This is in contrast to the later stage of planetary formation, where gravitational attraction plays a pivotal role independent of the  material composition.

Numerical simulation \citep{Wada} of dust aggregate collisions showed that the critical velocity for collisional growth, where the mass of the largest fragment after collision is larger than the mass of the larger object before collision, is  60\,m/s for H$_2$O ice  aggregates and they estimated 6\,m/s for silicate aggregates. This variation in the critical velocities is the result of differences in the surface energy and elasticity of the constituent materials. The shape of a dust grain is usually assumed to be spherical for simplicity. Previous numerical simulations \citep{Dominik, Wada} are based on theories \citep{Contact, Dominik95, Dominik96} that are applicable to two spherical grains. {\bf Moreover, the size of a composing grain has been assumed to be 0.1\,$\mu$m. If the size of a grain is smaller than 0.1\,$\mu$m, the critical velocity is higher than 60\,m/s (Wada, personal communication). On the other hand, a heating event quickly increases the size of icy grains \citep{kuroiwa}. The size of a grain is one of the key quantities to be investigated.}

If collision velocity of a dust grain (or aggregate) is lower than the critical velocity, dust grains and aggregates can grow. The collision velocity between dust aggregates is mainly determined by the turbulence of protoplanetary nebula gas \citep{Ormel} and radial migration of aggregates due to gas drag \citep{wei77}. A typical maximum collisional velocity is $\sim 50$\,m/s. This velocity is much larger than the critical velocity for silicate aggregates and roughly the same as that for icy aggregates. Based on these results, \cite{Okuzumi12} showed that an icy planetesimal can be formed through coagulation of H$_2$O ice dust aggregates without any special mechanism for {\bf concentrating} the aggregates. 

Based on the studies cited above, it seems that icy dust aggregates can grow sufficiently through collisional sticking. However, an important factor has been missing: H$_2$O ice is a volatile material. Its sublimation temperature is $\sim 160\,$K in protoplanetary nebula conditions. The location where the partial pressure of H$_2$O gas and the equilibrium vapor pressure of H$_2$O are equal is called the snow line. Inside this line, H$_2$O ice cannot exist. The vapor and solid phases of H$_2$O coexist outside this line. In this low-temperature region, sublimation and condensation of H$_2$O molecules proceed simultaneously. In equilibrium, both rates are the same.

To determine the location of the snow line, it is implicitly assumed that the surface of the solid phase is flat. However, the grain shape is assumed to be spherical. The curved surface of the solid phase induces Laplace pressure below the surface. Compressive stress arises below a convex surface and tensile stress arises below a concave surface. This stress slightly changes the chemical potential of the solid phase. As a result, the equilibrium vapor pressure on a convex surface is higher than that on a flat surface. The connection between two grains is called a ``neck'' because of its concave surface. On the main part of the grain (convex surface), the equilibrium vapor pressure is higher than that on the neck (concave surface). This difference results in transportation of an H$_2$O molecule from a convex to a concave surface, resulting in the growth of a neck. This process is called sintering \citep{Maeno, sintrev}.

The growth of a neck changes the mechanical interactions between grains. For example, smooth rolling that retains the connections between grains  is possible before sintering. This smooth rolling is enabled by the tensile stress at the edge of a neck and compressive stress around the center of a neck \citep{Contact}. The tensile stress around the neck edge appears because of cohesive contact between the two spherical grain surfaces. On the other hand, the center of the neck is mostly compressed to flatten the concave surface of the grain. A neck is covered by transported H$_2$O molecules after sintering. Then breakup of a neck is necessary for rolling motion because the tensile stress is not induced inside the grown part outside the original neck. If sintering does not take place, rolling friction dissipates a significant fraction of the initial kinetic energy of the aggregates \citep{Dominik}. We expect that sintering thus changes the collisional outcomes and affects the initial stage of planetary formation.

The collision of sintered dust aggregates was studied by \cite{Sirono}. The collisional outcome was bouncing because of the increased strength of the aggregate. The number of grains was $\sim$ 100 in that simulation. Based on that study, we increased the number of grains and investigated a wide range of parameter sets in the current simulation. The simulation was two-dimensional (2-D) in the sense that the motion of 3-D grain is limited on a plane as done by \cite{Dominik}. We determined dependence of collisional outcomes on the degree of sintering  and on the degree of compaction of colliding aggregates. The compaction degree changes because of collisions. Compacted aggregates are stronger than less-compacted aggregates, and the sintering degree changes the mechanical strength of an aggregate. These two factors should affect the collisional outcomes. 

In the next section, we describe how a neck grows. We model the mechanical interactions of a sintered contact in Section~3 and an outline of the numerical simulation is shown in Section~4. The numerical results are presented in Section~5, the effects on planetary formation are discussed in Section~6. Conclusions are given in Section~7.

\section{Growth of a neck}

\begin{figure}[h]
\begin{center}
\includegraphics*[width=3.5cm,height=11cm, angle=-90]{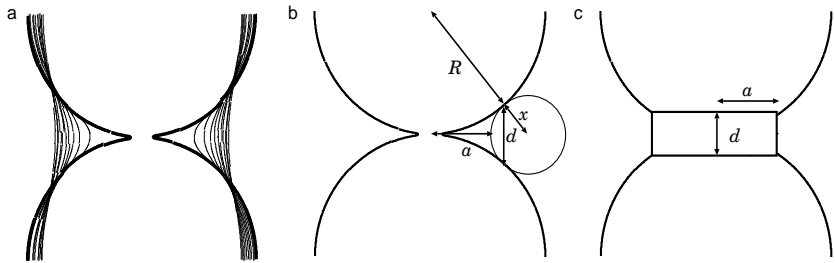}
\caption{(a) Evolution of the shape of a neck during sintering. The
 thick curve is the initial surface profile. The thin curves are the profiles at the same time interval of 0.02 in the normalized units \citep{sirono11}.  (b) The neck profile is approximated by a circle with a radius $x$ smoothly connected to spherical grains of radius $R$. (c) To calculate elastic responses, a neck is represented by a cylinder with a radius of $a$ and a thickness $d$.}
\label{fig:shape}
\end{center}
\end{figure}

A schematic of a neck between two spherical grains is shown in Figure~\ref{fig:shape} (thick curve). The radius of a neck without sintering depends on the surface energy, elasticity, and size of the grains \citep{Contact}. For two H$_2$O ice grains 0.1$\mu$m in size, the ratio $\beta=a/R$, where $a$ is the neck radius and $R$ is the grain radius, is 0.123. 

The molecules that comprise a grain migrate to the neck to minimize the total surface area. This process is called sintering. Many mechanisms contribute to the migration, including sublimation--recondensation, surface diffusion, grain boundary diffusion, and plastic deformation \citep{Maeno, sintrev}. The neck grows and the main part of the grain shrinks through migration. \cite{sirono11} calculated the growth rate of a neck assuming sublimation--recondensation. If other mechanisms contribute to the migration, the growth rate increases as compared to that reported by \cite{sirono11}. As shown in Figure~\ref{fig:shape}, if the equal-sized grains are  connected in a line to form a chain, the final shape of the chain is a cylinder with a radius of $\simeq 0.8R$. It can be seen that the growth rate of the neck is fast at the beginning and then decreases over time. This is because the migration rate of molecules is determined by the surface curvature of the neck: a deeply curved surface grows faster than less-curved  surface. 

\begin{figure}[h]
\begin{center}
\includegraphics*[width=12cm,height=8cm]{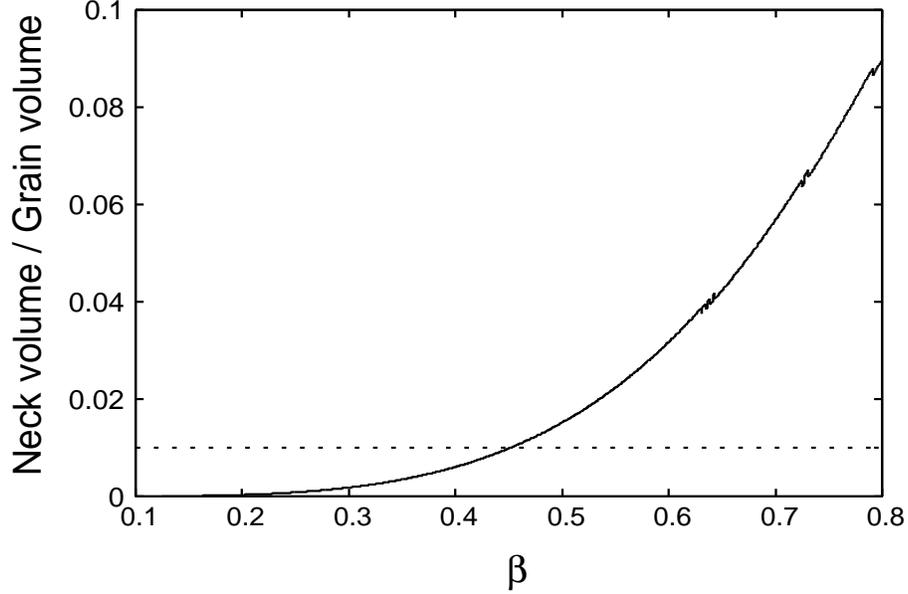}
\caption{Neck volume normalized by the grain volume as a function of $\beta=a/R$.}
\label{fig:vol}
\end{center}
\end{figure}

Figure~\ref{fig:vol} shows the increased evolution of the neck volume through the migration of molecules. The neck volume is defined as the space  between the original and evolved surfaces. It can be seen that significant sintering occurs   with only a small number of migrating molecules. The volume required for $\beta=0.5$ is only $\sim 2$\% of the volume of one grain. The main component of interstellar icy grains is H$_2$O, with other minor species such as CO, CO$_2$, NH$_3$, CH$_4$, CH$_3$OH, and C$_2$H$_2$. The concentrations of minor icy species are typically on the order of $\sim 1$\%, as shown by infrared observations \citep{Gibb}. Thus, sintering is caused not only by H$_2$O but also by the minor species. These species contribute to the formation of ``sintering zones'' at different heliocentric distances \citep{sirono11, okuzumi}. {\bf Physical properties of the minor species, such as surface energy and elasticity, are different from those of H$_2$O ice. In this study, we assume that the composing molecule is only H$_2$O for simplicity.}

\section{Interactions between grains}

\subsection{Sintered necks}

The shape of a pair of sintered grains is approximated by a circle of radius $x$ smoothly touching each sphere (Figure~\ref{fig:shape} ({\bf b})). The relation between the neck radius $a$ and $x$ is given by 
\begin{equation}
x={a^2\over 2(R-a)}.
\label{eq:a-x}
\end{equation}

\begin{figure}[h]
\begin{center}
\includegraphics*[width=6cm,height=7cm,angle=-90]{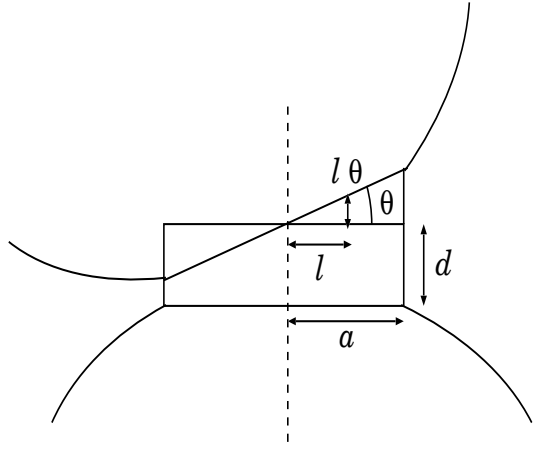}
\caption{A schematic figure of a rolling grain with a sintered contact. The upper grain tilts left with an angle of $\theta$. The radius and thickness of the neck is $a$ and $d$, respectively. Strain from the center with a distance of $l$ is given as $l\theta/d$.}
\label{fig:tilt}
\end{center}
\end{figure}

The thickness of the {\bf neck}, $d$, is the distance between two contact points on the grain surface (Figure~\ref{fig:shape}(b)), which is given by
\begin{equation}
d={2Rx\over R+x}.
\label{eq:d}
\end{equation}
Mechanical interactions between the two sintered  grains are approximated by those for an elastic disk (Figure~\ref{fig:shape}(c)) with a radius of $a$. It is assumed that only the disk is deformed elastically and that the other part of the grain is rigid. {\bf The mechanical interactions of this model are linear, and plastic deformation is not taken into account.} 

If the disk is stretched along the direction connecting the two grain centers by the amount of $u_{\rm ss}$, the stress $\sigma$ appearing in the disk is given by $\sigma=u_{\rm ss}E/d$, where $E$ is the Young's modulus of H$_2$O ice. Then, the total force required for stretching is $\pi a^2\sigma=\pi a^2 u_{\rm ss}E/d$. From this formula, the spring constant $k_{\rm ss}$ for stretching motion is given by
\begin{equation}
k_{\rm ss}={\pi a^2 E\over d}.
\label{eq:kss}
\end{equation}

If a force is applied parallel to the disk surface and a displacement of $u_{\rm st}$ is attained, the shear stress appeared in the disk is $\sigma=u_{\rm st}G/d$, where $G$ is the shear modulus. By the same argument for the stretching case, the spring constant  for tangential (sliding) motion, $k_{\rm st}$, is given by
\begin{equation}
k_{\rm st}={\pi a^2 G\over d}.
\label{eq:kst}
\end{equation}

Suppose the upper surface of the disk is tilted by amount of $\theta$ because of an external moment $M$ with respect to the axis on the center of the disk surface. In this case, the net vertical force is zero. At a distance $l$ from the axis, the vertical strain is given as $\theta l/d$ (Figure~\ref{fig:tilt}). The stress accompanied by this strain is $\theta lE/d$. By integrating this stress throughout the surface, the moment $M$ around the axis on the surface is obtained as
 \begin{eqnarray}
M&=&2\int^a_{-a} {\theta lE\over d}\sqrt{a^2-l^2}ldl\nonumber\\
&=&{\pi a^4E\over 4d}\theta=k_{\rm sr}\theta.
\label{eq:M}
\end{eqnarray}
The elastic energy stored is written as $k_{\rm sr}\theta^2/2$.

The neck is broken when a large force is applied. Fracturing takes place when the stress inside a solid material exceeds its tensile strength, which strongly depends on the number and size of internal cracks. A crack grows by breaking interatomic or intermolecular bonds at the crack edge. The elastic energy stored around a crack edge is consumed to break the bonds. When torque is applied to rotate a grain relative to an adjacent grain, the stress reaches tensile strength $T$ at the edge of a neck when $\theta aE/d=T$. To complete fracturing, the elastic energy stored should be larger than the surface energy, $\pi a^2 \gamma$. If the elastic energy is much larger than the surface energy, cracks propagate quickly and the material breaks at once. If the elastic energy is small, fragmentation stops after breaking some bonds. In this case, the elastic energy is insufficient to break the neck completely at the beginning of fracture. {\bf For simplicity, instantaneous breaking of the neck is assumed in this study. This point should be addressed by laboratory experiments.}

{\bf Under the assumption that the neck fully breaks at once,} the neck is broken when the displacement $u$ ($\theta$ for rolling) reaches $k_{i}u^2/2=\pi a^2 \gamma$ ($i$ is rolling, stretching, or sliding). From this condition, critical displacements for stretching $u_{\rm ss,c}$, for sliding $u_{\rm st,c}$ and for rolling $\theta_{\rm c}$, are respectively given by
\begin{equation}
u_{\rm ss,c}=\sqrt{2\pi a^2\gamma\over k_{\rm ss}},\quad 
u_{\rm st,c}=\sqrt{2\pi a^2\gamma\over k_{\rm st}},\quad
\theta_{\rm c}=\sqrt{2\pi a^2\gamma\over k_{\rm sr}}.
\label{eq:Ps2}
\end{equation}

\subsection{Non-sintered necks}

When two sintered aggregates collide, some of the necks might break and new contacts between grains might form. Because the temperature increase associated with a collision is very low, a newly formed contact is a non-sintered one. The mechanical responses of a non-sintered necks were described by \cite{Contact, Dominik}. We adopted the same equations used in those studies. {\bf Recently, a model taking account of adhesion hysteresis \citep{Krijt13, Krijt14} has been developed. A simulation using the model is desirable to determine the effect of adhesion hysteresis for non-sintered aggregates.} 

The radius of a non-sintered neck at the equilibrium is 
\begin{equation}
a_0=\left({9\pi \gamma R^{*2}\over E^*}\right)^{1/3},
\label{eq:Fss}
\end{equation}
where $E^*=E/2(1-\nu^2)$ ($\nu$ is Poisson's ratio). It should be noted that $R^*$ is the reduced radius of two contact grains, given by $R_1R_2/(R_1+R_2)$, with radii of $R_1$ and $R_2$. In this study, we focus on a monodispsrsed system and $R^*=R/2$. 

When a force $F$ is applied along the stretching direction, the neck radius changes. The relation between $F$ and $a$ can be written by
\begin{equation}
{F\over F_{\rm c}}=4\left[\left({a\over a_0}\right)^3-\left({a\over a_0}\right)^{3/2}\right],
\label{eq:fa}
\end{equation}
where $F_{\rm c}=3\pi \gamma R^*$ is the critical force required to break a non-sintered contact. On the other hand, the compression length (the shift of the distance between two centers of spheres) $\delta$ can be written as a function of $a$ as
\begin{equation}
{\delta \over \delta_0}=3\left({a\over a_0}\right)^2-\left({a\over a_0}\right)^{1/2}.
\label{eq:deltaa}
\end{equation}
Combining Equations~(\ref{eq:fa}) and (\ref{eq:deltaa}), we can obtain the relation between $F$ and $\delta$ implicitly.

For rolling and sliding motions, simple elastic responses are assumed until the forces reach the critical values. The elastic energy stored as rolling deformation, $E_{\rm nr}$, is written using the spring constant, $k_{\rm nr}$, for rolling motion as
\begin{equation}
E_{\rm nr}={1\over 2}k_{\rm nr}\xi^2={1\over 2}{4F_{\rm c}\over R^*}\xi^2,
\label{eq:kr}
\end{equation}
where the rolling displacement is $\xi=R\theta/2$ (see Figure~2(c) of \cite{Wada07}). When $\xi$ reaches  $\xi_{\rm c}$, the spring is cut and the stored elastic energy dissipates. This corresponds to the rolling friction due to the breaking of intermolecular bonds. In the simulation, the spring was newly formed immediately after the breaking. The estimated values of $\xi_{\rm c}$ varied from 2\,\AA\citep{Dominik95} to 32\,\AA\,\citep{Heim}. We adopted $\xi_{\rm c}=2$\,\AA\, and 30\,\AA\, in the simulation.

The critical force for tangential motion, $F_{\rm nt}$, is given by
\begin{equation}
F_{\rm nt}={Ga_0^2\over 2\pi},
\label{eq:F}
\end{equation}
and this spring is broken when the displacement reaches
\begin{equation}
u_{\rm nt}={2-\nu \over 16\pi}a_0.
\label{eq:u}
\end{equation}
As with the rolling motion, the spring immediately reformed after breaking.

\subsection{Shrinkage of a grain}

As shown in Figure~\ref{fig:shape}(a), the main part of a spherical grain shrinks during sintering. A sintered contact is broken mainly through rolling motion, as shown later. If the degree of sintering is low, the main part of the grain retains the original shape. In this case, a non-sintered contact is likely to form because the distance between the centers of two grains is slightly shorter than the sum of the two radii ($-4.2\times 10^{-4}\,\mu$m for two 0.1$\mu$m radius spherical grains). On the other hand, if the degree of sintering is high, the radius decreases substantially ($-0.091\,\mu$m for a 0.1$\mu$m radius spherical grain when $\beta$ increases to 0.7 {\bf provided that the radius of grains is uniform}) {\bf The shrinkage depends on the size distribution of grains, which is determined by the thermal history of the grains \citep{kuroiwa}. This point should be addressed separately from this study.} 

\begin{figure}[h]
\begin{center}
\includegraphics*[width=12cm,height=8cm]{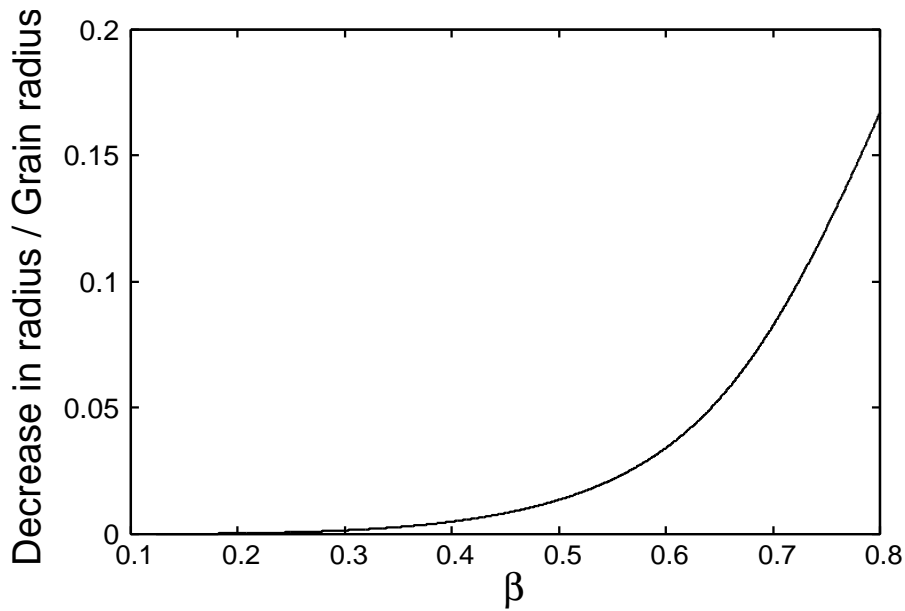}
\caption{Shrinkage of grain radius. The amount of shrinkage normalized by the initial grain radius as a function of $\beta$. When $\beta=0.7$, the decrease in radius is 0.09.}
\label{fig:shrink}
\end{center}
\end{figure}

This shrinkage should affect the formation of a new contact immediately after the breaking of a sintered neck. In this study, the grain radius $R$ decreased to $R'$, as shown in Figure~\ref{fig:shrink} during sintering. A new contact between sintered grains forms when the distance between the two grains is less than $2R'$. The radius of the contact area, $a_0$, and other quantities in Section~2.2 were calculated using the decreased $R'$. It should be noted that Figure~\ref{fig:shrink} is the result for a straight chain of spherical grains. If the number of contacts per grain is more than two, the decrease in $R$ should be larger than that shown in Figure~\ref{fig:shrink}. When taking shrinkage into account, a new contact formation is less probable than in the case without shrinkage.

\section{Numerical simulation settings}

We simulated aggregates composed of H$_2$O ice. The material parameters were adopted from \cite{Dominik}: Young's modulus $E=7.0\times 10^{9}\,{\rm Pa}$, shear modulus $G=2.8\times 10^{9}\,{\rm Pa}$, Poisson's ratio $\nu=0.25$, surface energy $\gamma=0.10\,{\rm J\,m^{-2}}$, and a critical distance for rolling motion $\xi_{\rm c}=2$\,\AA\, as a standard value. For comparison, $\xi_{\rm c}=30$\,\AA\, cases were also simulated. The initial grain radius $R$ was 0.1$\mu$m. We simulated head-on collisions of two equal-mass aggregates. The number of grains in each aggregate was 1024, which is quite small. A large-scale simulation with 3-D aggregates will be presented in a separate paper. 

\begin{figure}[h]
\begin{center}
\includegraphics*[width=12cm,height=4cm]{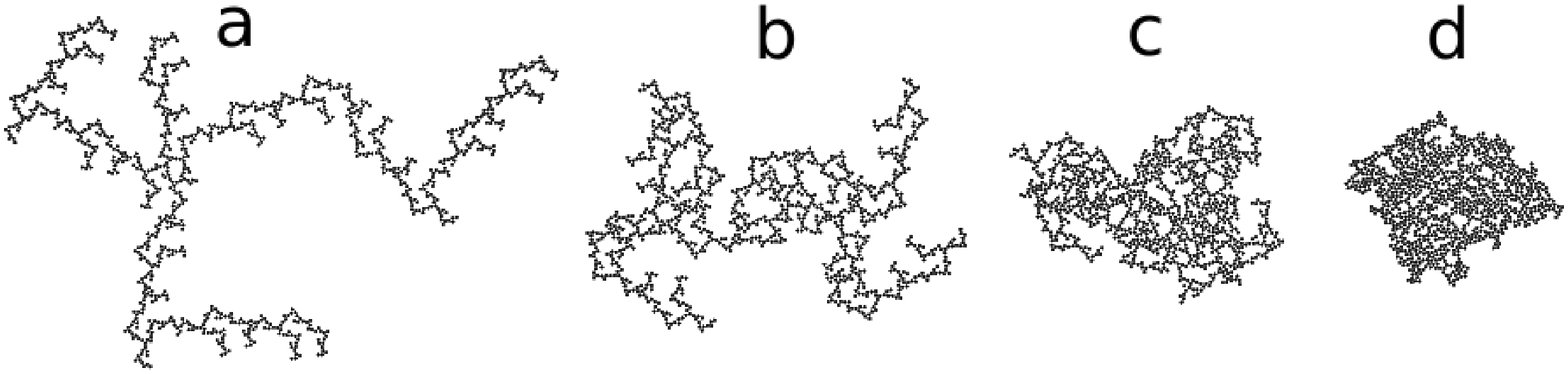}
\caption{The initial arrangement of grains: (a) BCCA aggregate, compacted aggregates with growth velocities of (b) 1, (c) 3, and (d) $20\,{\rm m\,s}^{-1}$, respectively.}
\label{fig:init}
\end{center}
\end{figure}

The arrangements of grains in an aggregate is determined using one of two methods. One is the so-called ballistic-cluster-cluster-aggregation method (BCCA, \cite{Meakin}). Initially, two grains collide to form an aggregate. The motion of each grain stops immediately after making a new contact. In the next step, two aggregates of two grains collide to form an aggregate of four grains. Before collision, {\bf the orientation of aggregates is given randomly, and only linear motion is given  without rotation.} The motion of the new aggregate again stops when a new contact is formed. An aggregate of 1024 grains is eventually formed by continuing this sequence. We produced five BCCA aggregates by changing random numbers that determine the rotation. An example of a BCCA aggregate is shown in Figure~\ref{fig:init}(a). It has an extremely porous structure. The basic unit of the aggregate is a one-dimensional chain of grains.  

In the BCCA procedure, the motion of aggregates stops when a new contact is formed. Actually, the motions of the grains remain after a new contact is formed.  In this case, the arrangement of grains depends not only on the initial orientations of the aggregates but also on the collision velocity. We performed collision simulations to prepare aggregates consisting of 1024 grains. The collision velocity used in this simulation was defined as the growth velocity $V_{\rm g}$. The relative motions of the grains were computed based on the interactions without sintering. Examples of aggregates produced by this procedure are shown in Figures~\ref{fig:init}(b) ($V_{\rm g}=1\,{\rm m\,s}^{-1}$), (c) ($V_{\rm g}=3\,{\rm m\,s}^{-1}$), and (d) ($V_{\rm g}=20\,{\rm m\,s}^{-1}$).

\begin{figure}[h]
\begin{center}
\includegraphics*[width=12cm,height=8cm]{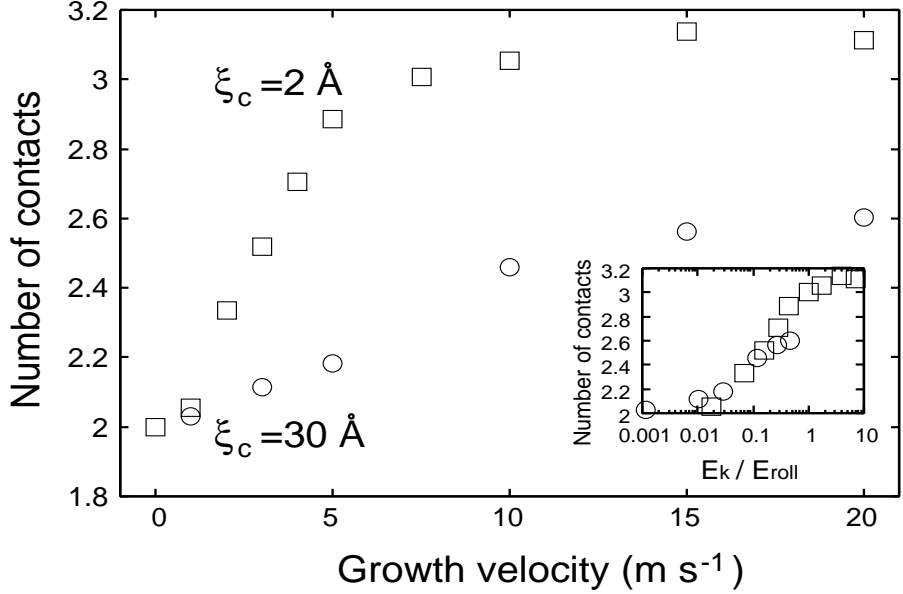}
\caption{Average number of contacts on a grain as a function of growth velocity. Squares and circles are the data for the critical displacement of rolling motion $\xi_{\rm c}=2$\,\AA\, and 30\,\AA, respectively. Inset: same as the main panel, but the horizontal axis is the ratio between the initial kinetic energy $E_{\rm k}$ to the energy $E_{\rm roll}$ required to roll a grain by 90$^\circ$ around the adjacent grain.}
\label{fig:initn}
\end{center}
\end{figure}

We conducted simulations using aggregates produced with growth velocities of $V_{\rm g}=1$, 2, 3, 4, 5, 7.5, 10, 15, and $20\,{\rm m\,s}^{-1}$ for $\xi_{\rm c}=2$\,\AA. For comparison, $V_{\rm g}=1$, 3, 5, 10, 15 and $20\,{\rm m\,s}^{-1}$ cases were simulated for $\xi_{\rm c}=30$\,\AA. The structure of an aggregate changed as $V_{\rm g}$ increased. When $V_{\rm g}$ was large, an aggregate became compact. Figure~\ref{fig:initn} shows the average number of contacts per grain as a function of $V_{\rm g}$. Zero collision velocity corresponds to a BCCA aggregate. The number of contacts is approximately two at low $V_{\rm g}$ and three at high $V_{\rm g}$. At low $V_{\rm g}$, the number is two because the basic structure is a chain of grains. At high $V_{\rm g}$, the number is three because that is the maximum number of contact points if only rolling motion is possible \citep{sirono00}. This is a good approximation for non-sintered contacts because sliding and stretching motion require larger forces than rolling motion. For high-velocity collisions such that $V_{\rm g}>20\,{\rm m\,s}^{-1}$, the induced stresses during a collision are high enough for stretching and sliding to exceed the contact number of three. 

In the inset of Figure~\ref{fig:initn}, the horizontal axis shows the kinetic energy per grain, $E_{\rm k}=mV_{\rm g}^2/4$ ($m$: mass of a grain), normalized by the energy required to roll a grain by 90$^\circ$ around an adjacent grain, $E_{\rm roll}=6\pi^2\gamma R\xi_{\rm c}$. It can be seen that {\bf both data sets fall on the same curve}. This figure shows that the ratio $E_{\rm k}/E_{\rm roll}$ determines the number of contacts.

We conducted collision simulations using the BCCA aggregates and compacted aggregates described above. The degree of sintering represented by $\beta$ was varied as $\beta=0.123$ (without sintering for comparison), 0.2, 0.3, 0.4, 0.5, 0.6, and 0.7. The collision velocity $V_{\rm c}$ was varied between 1 and 100$\,{\rm m\,s}^{-1}$.

\section{Numerical results}

\subsection{BCCA aggregates}

\begin{figure}[h]
\begin{center}
\includegraphics*[width=14cm,height=2.5cm]{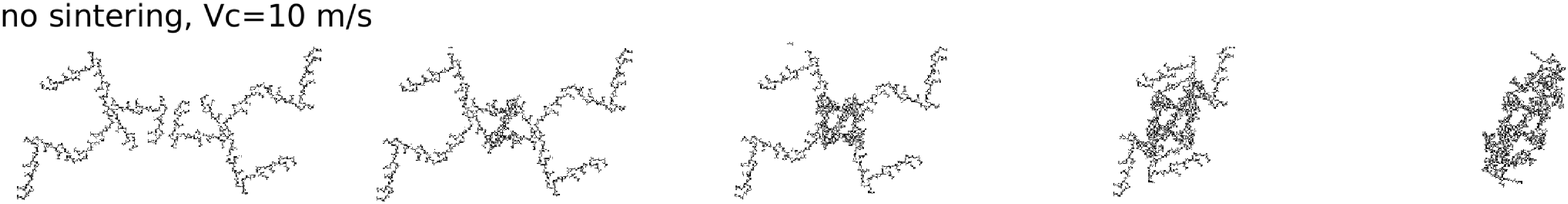}
\includegraphics*[width=14cm,height=2.5cm]{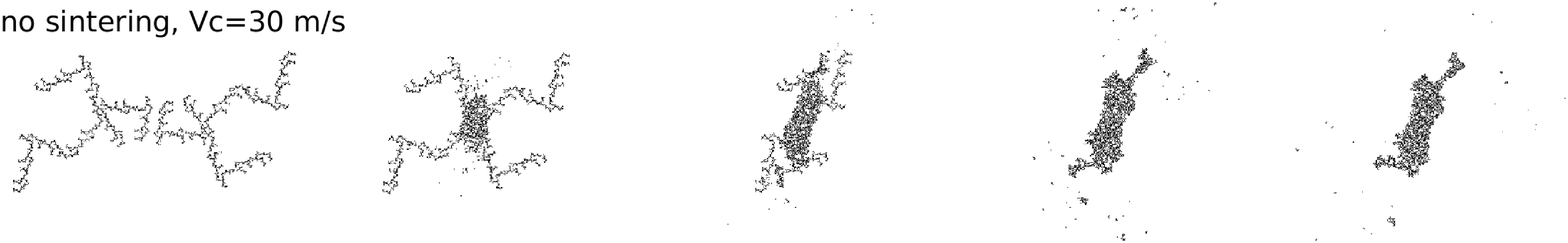}
\includegraphics*[width=14cm,height=2.5cm]{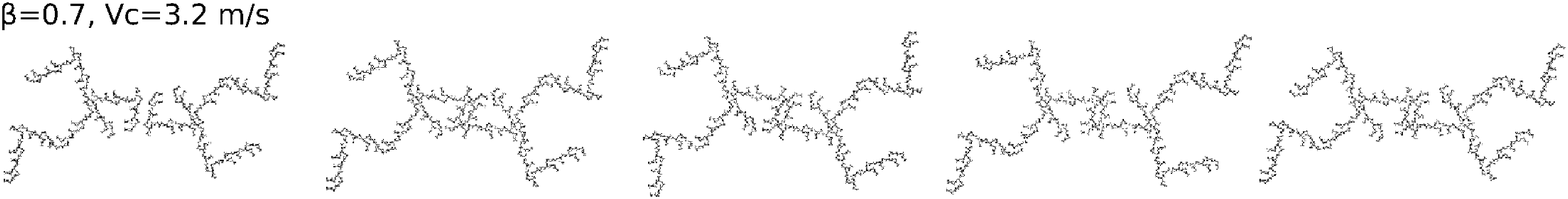}
\includegraphics*[width=14cm,height=2.5cm]{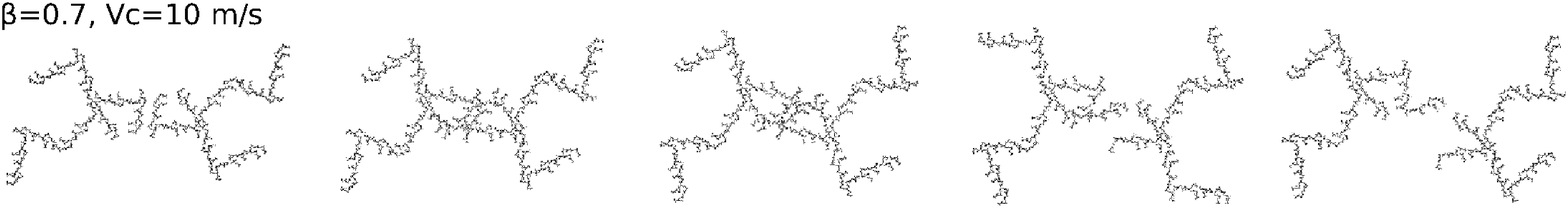}
\includegraphics*[width=14cm,height=2.5cm]{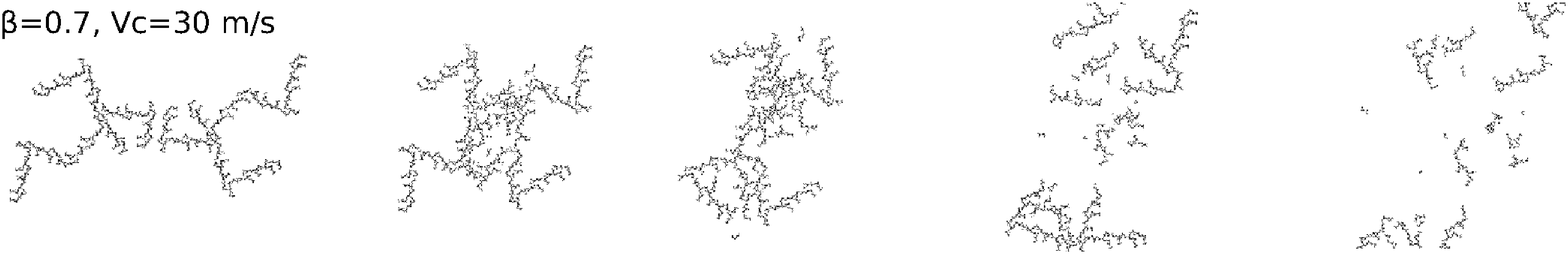}
\caption{{\bf Snapshots of collisions of two BCCA aggregates. From top to bottom, $V_{\rm c}=10\,{\rm m\,s^{-1}}$ without sintering, $30\,{\rm m\,s^{-1}}$ without sintering, $3.2\,{\rm m\,s^{-1}}$ with sintering ($\beta=0.7$), $10\,{\rm m\,s^{-1}}$ with sintering ($\beta=0.7$), and $30\,{\rm m\,s^{-1}}$ with sintering ($\beta=0.7$). From left to right, snapshots at times of 0, 2500, 5000, 10000, and 20000 in the normalized time units.} The time is normalized with the vibration period of stretching motion for a non-sintered contact.}
\label{fig:coltime}
\end{center}
\end{figure}

Typical examples of collisional outcomes of BCCA aggregates are shown in Figure~\ref{fig:coltime}. Non-sintered BCCA aggregates stick perfectly  at a collision velocity of $V_{\rm c}=10\,{\rm m\,s^{-1}}$. Even at a higher velocity of $V_{\rm c}=30\,{\rm m\,s^{-1}}$, the sticking efficiency (defined as the ratio of maximum aggregate mass after collision to the total mass) is close to unity.

On the other hand, {\bf three panels from the bottom in} Figure~\ref{fig:coltime} show the typical examples of the collisional outcomes of sintered aggregates. Two sintered aggregates stick at $V_{\rm c}=3.2\,{\rm m\,s^{-1}}$, bounce at  $V_{\rm c}=10\,{\rm m\,s^{-1}}$, and totally rupture at  $V_{\rm c}=30\,{\rm m\,s^{-1}}$. Except at low velocities, these collisional outcomes differ substantially from those without sintering. Because sintering increases the elasticity  and the strength of the neck between grains, energy dissipation through rolling motion cannot proceed. This is in contrast to the non-sintered cases, where rolling friction dissipates kinetic energy. As a result, bouncing is observed at $V_{\rm c}=10\,{\rm m\,s^{-1}}$. For low collisional velocities, two sintered aggregates stick together without any notable deformation. At high collision velocities, the aggregates extensively disrupt to small fragments. 

It might be considered that the high energy required to break a sintered contact leads to more efficient sticking as compared to non-sintered aggregates because of higher energy dissipation. This is not the case because only a limited number of contacts where stress is concentrated can break. Most of the contacts remain intact and do not contribute to energy dissipation. This is common phenomena observed in the breaking of solid material. Brittle disruption occurs through propagation of cracks, and the breaking of intermolecular bonds occurs only at the edges of cracks where stress is concentrated. Thus, a substantial fraction of the initial kinetic energy remains even after the collision.

This is a totally different deformation mode from that in non-sintered aggregates. In a collision of non-sintered aggregates, the aggregate deforms through rolling of the grains. Because the critical force to  roll is quite small, rolling deformation occurs throughout an aggregate. This deformation mode is similar to that of liquid, where molecules smoothly rotate around adjacent ones.

\begin{figure}[h]
\begin{center}
\includegraphics*[width=6cm,height=16cm,angle=-90]{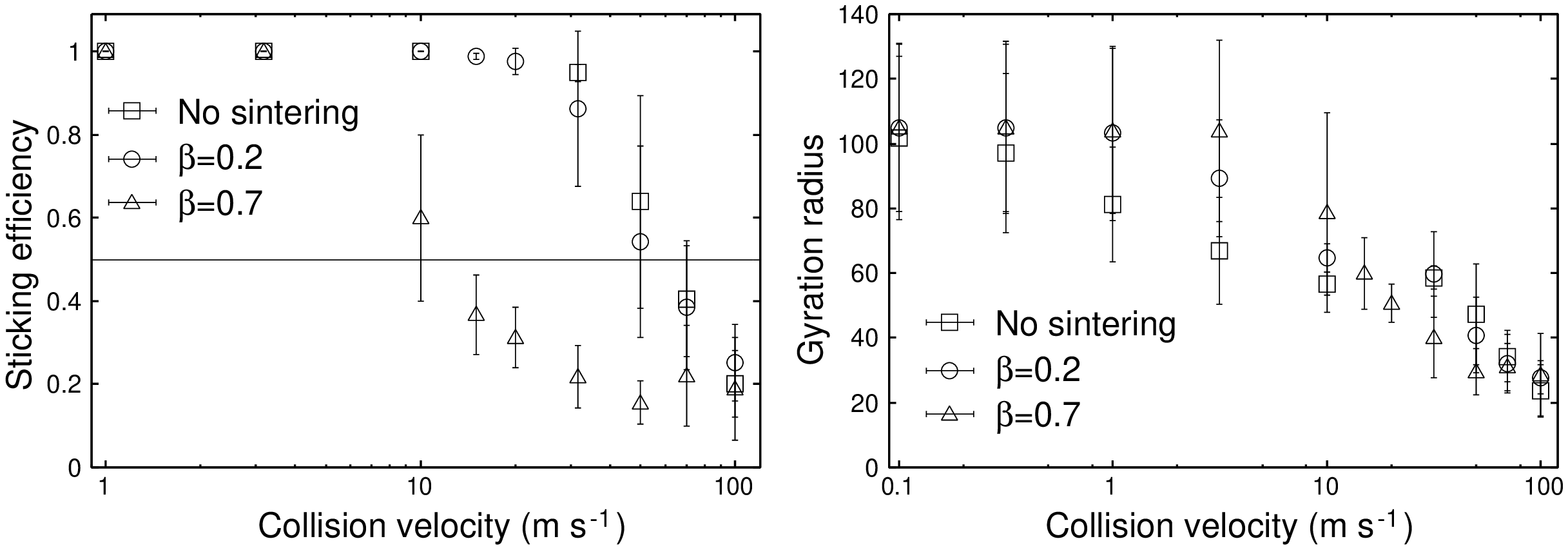}
\caption{{\bf Left:} sticking efficiency of BCCA aggregate.  {\bf Right: gyration radius normalized by grain radius as a function of collision velocity. In both panels, squares, circles, and triangles represent non-sintered, and sintered with $\beta=0.2$ and 0.7, respectively.  Error bars are the standard deviation of five runs using different arrangements of grains.}}
\label{fig:BCCA-s}
\end{center}
\end{figure}

Figure~\ref{fig:BCCA-s}~left shows the sticking efficiency, defined as the ratio of the maximum mass after collision to the total mass, as a function of collision velocity. It can be seen that non-sintered aggregates can grow (sticking efficiency more than 0.5) at $V_{\rm c}=50\,{\rm m\,s^{-1}}$. This is consistent with previous studies \citep{Wada07, Wada}. On the other hand, substantial disruption is observed even at $V_{\rm c}=15\,{\rm m\,s^{-1}}$ for the $\beta=0.7$ sintered aggregates. A $\beta=0.7$ aggregate bounces at $V_{\rm c}=10\,{\rm m\,s^{-1}}$. For velocities lower than $V_{\rm c}=3.2\,{\rm m\,s^{-1}}$, the $\beta=0.7$ aggregate sticks. 

The collisional outcomes of $\beta=0.2$ aggregates are the almost same as those of non-sintered aggregates. This is because the neck is not strong enough to sustain the stress induced during a collision. Moreover, if a sintered neck is broken, a new neck without sintering is  formed because the shrinkage of a grain does not proceed at low $\beta$ (see Figure~\ref{fig:shrink}).

Figure~\ref{fig:BCCA-s}~right shows the gyration radius of the largest aggregate after collision. {\bf The gyration radius is a measure of the aggregate size defined by}
\begin{equation}
R_{\rm g}=\sqrt{{1\over N}\sum_i (\vec{r}_i-\vec{r}_0)^2},
\end{equation}
{\bf where $\vec{r}-\vec{r_0}$ is the positional vector measured from the center of mass $\vec{r}_0$.} The gyration radius decreases monotonically as the collision velocity increases because of compaction and fragmentation. Below $10\,{\rm m\,s^{-1}}$, the gyration radius of the non-sintered aggregate is the smallest among the three types of aggregates. This shows that compaction through the rolling motion of the grains proceeds efficiently  in non-sintered aggregates. Compaction does not proceed in the $\beta=0.7$ aggregate below $10\,{\rm m\,s^{-1}}$. The evolution of porosity \citep{kataoka} would thus be affected by sintering. The sintered aggregate has much less density than the non-sintered aggregate. Above $10\,{\rm m\,s^{-1}}$, the gyration radius of $\beta=0.7$ aggregate is the smallest as compared to the other types of aggregates because of fragmentation.

\subsection{Compacted aggregates}

\begin{figure}[h]
\begin{center}
\includegraphics*[width=14cm,height=3cm]{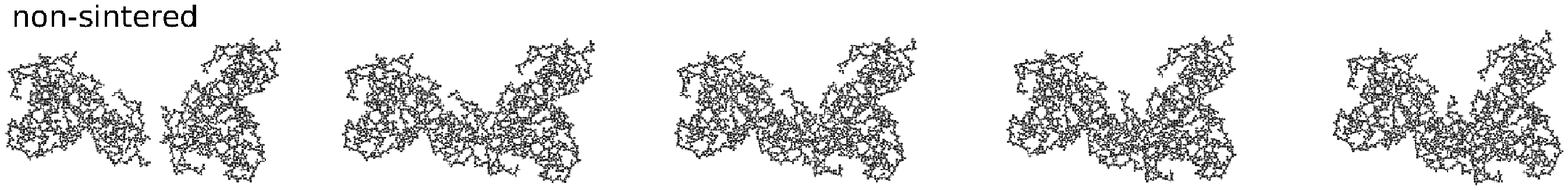}
\includegraphics*[width=14cm,height=3cm]{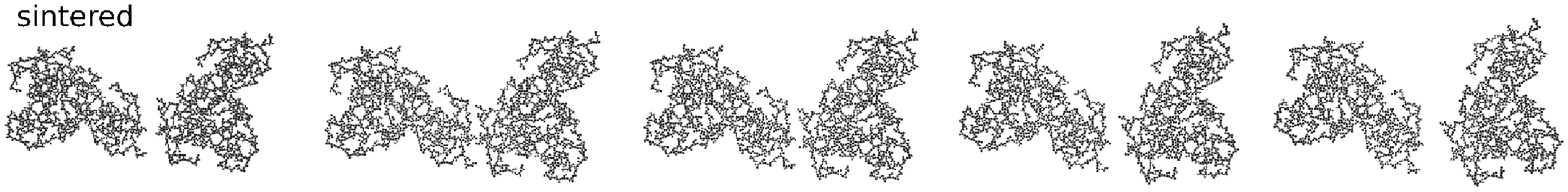}
\caption{Snapshots of collisions between two compacted aggregate with a growth velocity of $3\,{\rm m\,s^{-1}}$. Collision velocity is $V_{\rm c}=3\,{\rm m\,s^{-1}}$. Top: non-sintered aggregate. Bottom: sintered aggregate with $\beta=0.7$. From left to right, snapshots taken at times of 0,  625,  1250,  2500, and 5000 in the normalized time units. The time is normalized with the vibration period of stretching motion for a non-sintered contact.}
\label{fig:compact}
\end{center}
\end{figure}

Next, we conducted the collision simulations using compacted aggregates with $\beta=0.7$. Snapshots during collisions are shown in Figure~\ref{fig:compact}. The aggregate grown with $V_{\rm g}=3\,{\rm m\,s^{-1}}$ has a compacted structure as compared to the BCCA aggregate. Then, the compacted aggregates collided with a collision velocity of $V_{\rm c}=3\,{\rm m\,s}^{-1}$. The collisional outcome for the non-sintered case is shown in Figure~\ref{fig:compact}, top. Even when an aggregate is compacted, the non-sintered aggregates stick perfectly. On the other hand, the sintered aggregates cannot stick; instead, they bounce. This is simply because a compacted aggregate is stiffer than a BCCA aggregate. For BCCA aggregates, perfect sticking occurs when $V_{\rm c}\le 3.2\,{\rm m\,s^{-1}}$ irrespective of the sintering degree (see Figure~\ref{fig:BCCA-s}~left).

\begin{figure}[h]
\begin{center}
\includegraphics*[width=12cm,height=8cm]{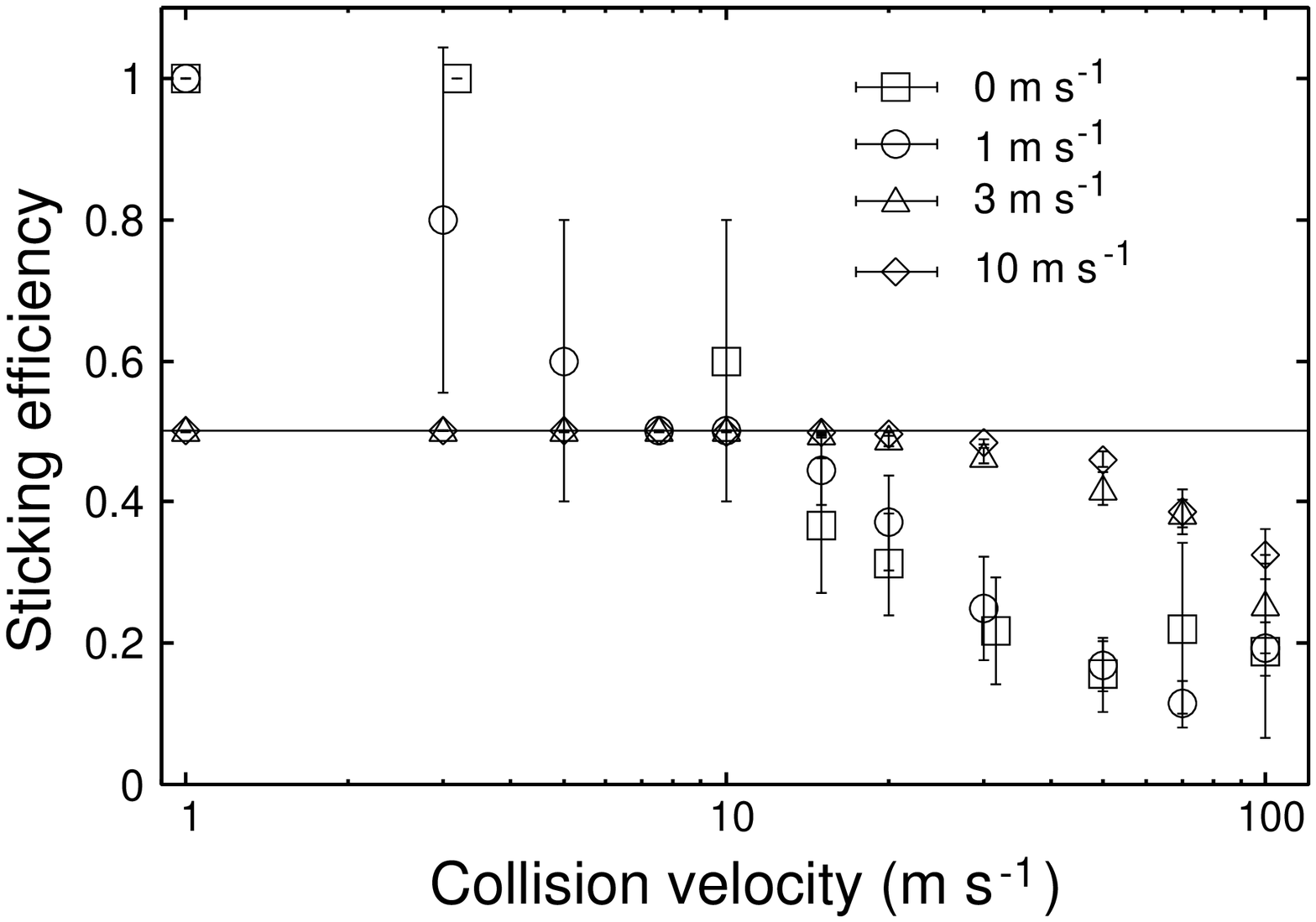}
\caption{Sticking efficiencies for sintered aggregates with various compaction degrees varied by growth velocity $V_{\rm g}$. Squares: $V_{\rm g}=0\,{\rm m\,s}^{-1}$ (BCCA aggregate), circles: $1\,{\rm m\,s}^{-1}$, triangles: $3\,{\rm m\,s}^{-1}$, diamonds: $10\,{\rm m\,s}^{-1}$. Error bars are the standard deviation of five runs with five different arrangement of grains (squares) and five different orientations of aggregates (circles, triangles, and diamonds).}
\label{fig:comp-s}
\end{center}
\end{figure}

Figure~\ref{fig:comp-s} compares the sticking efficiencies for compacted aggregates as a function of collision velocity. Data for $0\,{\rm m\,s^{-1}}$ correspond to BCCA aggregates with $\beta=0.7$; the results for {\bf $0\,{\rm m\,s^{-1}}$} are identical to {\bf (triangles)} shown in Figure~\ref{fig:BCCA-s}~left. When the growth velocity is low (0 and $1\,{\rm m\,s^{-1}}$), perfect sticking is realized at collision velocities less than $3\,{\rm m\,s^{-1}}$. Above $10\,{\rm m\,s^{-1}}$, substantial fragmentation takes place, and sticking efficiency is approximately 0.2. 

For aggregates with growth velocities of $3\,{\rm m\,s^{-1}}$ and $10\,{\rm m\,s^{-1}}$, the collisional outcome is  bouncing. Above $30\,{\rm m\,s^{-1}}$, fragmentation occurs.

%It should be reminded that the compacted aggregates are produced using non-sintered aggregates. A non-sintered aggregate compacts during a collision as shown in Fig.~\ref{fig:coltime}. In a protoplanetary nebula, aggregates drift to the central star due to gas drag. Growth is possible even an aggregate is compacted if sintering does not take place, as shown in Fig.~\ref{fig:compact}. However, if a non-sintered aggregate drifts into a sintering region, it cannot grow on a collision but bounces. The results show in Fig.~\ref{fig:comp-s} suggest that the growth of an aggregate is possible only when collision velocity is less than $\sim 100\,{\rm cm\,s^{-1}}$. This condition limits the maximum size of an aggregate infalling from outer non-sintered region. An aggregate grown with $100\,{\rm cm\,s^{-1}}$ can grow further even in a sintering region. However, if an aggregate grown with collision velocity of $300\,{\rm cm\,s^{-1}}$, it cannot grow in a sintering region.  

\subsection{Sintering degree dependence}

\begin{figure}[h]
\begin{center}
\includegraphics*[width=6cm,height=16cm,angle=-90]{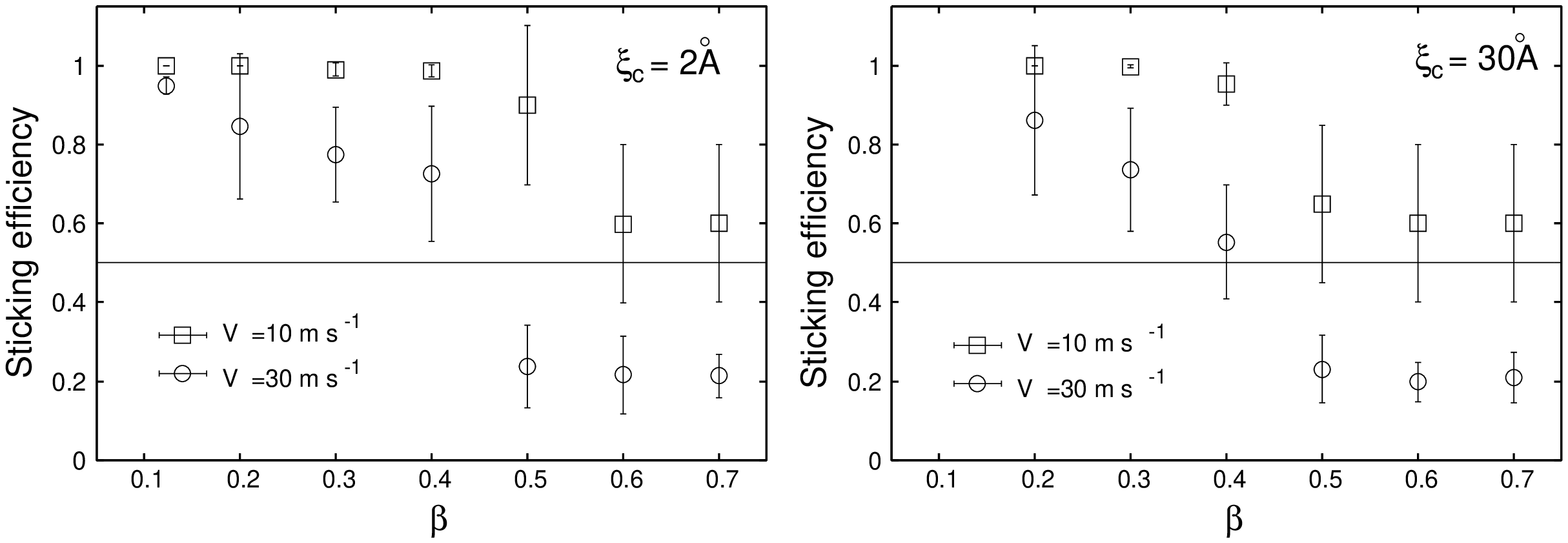}
\caption{Sticking efficiencies as a function of $\beta$ for BCCA aggregates with {\bf $\xi_{\rm c}=2$\AA (left) and $\xi_{\rm c}=30$\AA (right)}, with collision velocities of $V_{\rm c}=10\,{\rm m\,s^{-1}}$ (squares) and $V_{\rm c}=30\,{\rm m\,s^{-1}}$ (circles). Error bars are the standard deviations for five runs using five different aggregates.}
\label{fig:v10v30beta-s}
\end{center}
\end{figure}

We have shown the results of sintered BCCA aggregate collisions with $\beta=0.2$ and 0.7. We also investigated the $\beta$ dependence of collisional outcomes. Figure~\ref{fig:v10v30beta-s} shows the sticking efficiency of collisions with BCCA aggregates of various sintering degrees as a function of $\beta$ {\bf with $\xi_{\rm c}=2$\,\AA (left) and 30\,\AA (right)}. The collisional velocities were 10 and $30\,{\rm m\,s^{-1}}$. It can be seen that the sticking efficiency decreases as $\beta$ increases in both cases. If the collisional velocity is $10\,{\rm m\,s^{-1}}$, growth is possible for all values of $\beta$. This has been already suggested in Figure~\ref{fig:BCCA-s}~left, where the sticking efficiency of the $\beta=0.7$ aggregate with a collision velocity of $10\,{\rm m\,s^{-1}}$ is larger than unity. 

On the other hand,  qualitative changes were observed for $V_{\rm c}=30\,{\rm m\,s^{-1}}$ collisions. Collisional growth took place only for then $\beta\le 0.4$ aggregates. Above $\beta=0.5$, bouncing and  fragmentation occurred, and collisional growth was impossible. It should be noted that the sticking efficiency decreased between $\beta=0.4$ and 0.5 for both collision velocities. For the BCCA aggregates, the  sintering degree of $\beta=0.5$ was critical in the sense that the collisional growth was affected above this value. It should be noted that we have shown that the volume required for $\beta=0.5$ is 2\% of the volume of a grain (Figure~\ref{fig:vol}).

\begin{figure}[h]
\begin{center}
\includegraphics*[width=12cm,height=8cm]{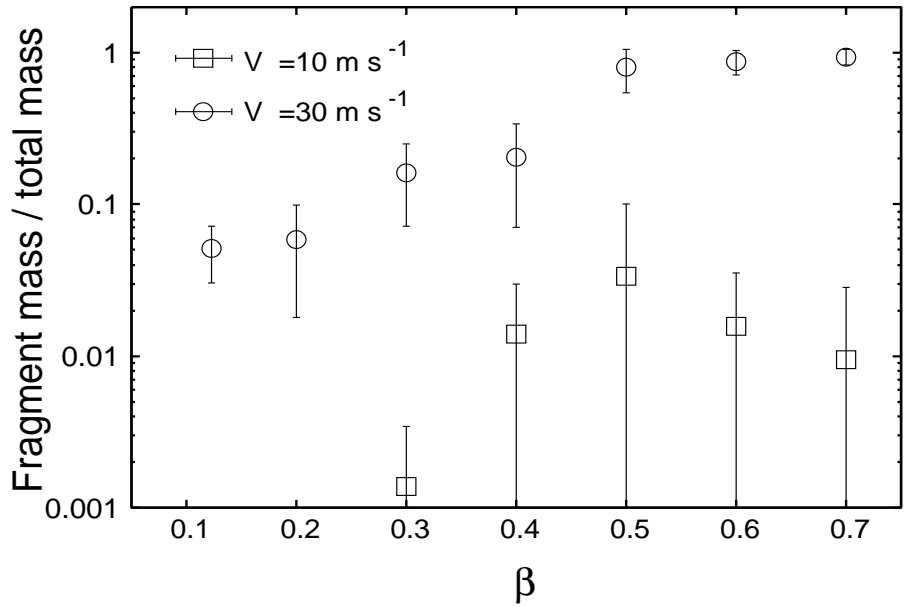}
\caption{{\bf The fraction of mass in fragments compared to the total mass} as a function of $\beta$  produced during collisions of BCCA aggregates with $V_{\rm c}=10\,{\rm m\,s^{-1}}$ (squares) and $V_{\rm c}=30\,{\rm m\,s^{-1}}$ (circles). {\bf A fragment is defined as an aggregate whose mass is less than one-fourth of the total mass.} Error bars are the standard deviations for five runs using five different aggregates.}
\label{fig:v10v30beta-f}
\end{center}
\end{figure}

The {\bf mass} of fragments produced during a collision is presented in Figure~\ref{fig:v10v30beta-f}. A fragment is defined as an aggregate whose mass is less than one-fourth of the total mass. The fragment mass of $V_{\rm c}=30\,{\rm m\,s^{-1}}$ collisions monotonically increases as $\beta$ increases. On the other hand, the fragment mass of $V_{\rm c}=10\,{\rm m\,s^{-1}}$ collisions has a peak at $\beta=0.5$. Aggregates of $\beta=0.7$ produced only a small mass of fragments. This is because an aggregate is hard, and thus few  necks were broken. On the other hand, if $\beta$ is low, an aggregate is too weak to sustain the stress, and plastic deformation through rolling occurs as in non-sintered aggregates. In these cases, almost perfect sticking takes place and the amount of fragments is negligible.

\begin{figure}[h]
\begin{center}
\includegraphics*[width=6cm,height=16cm,angle=-90]{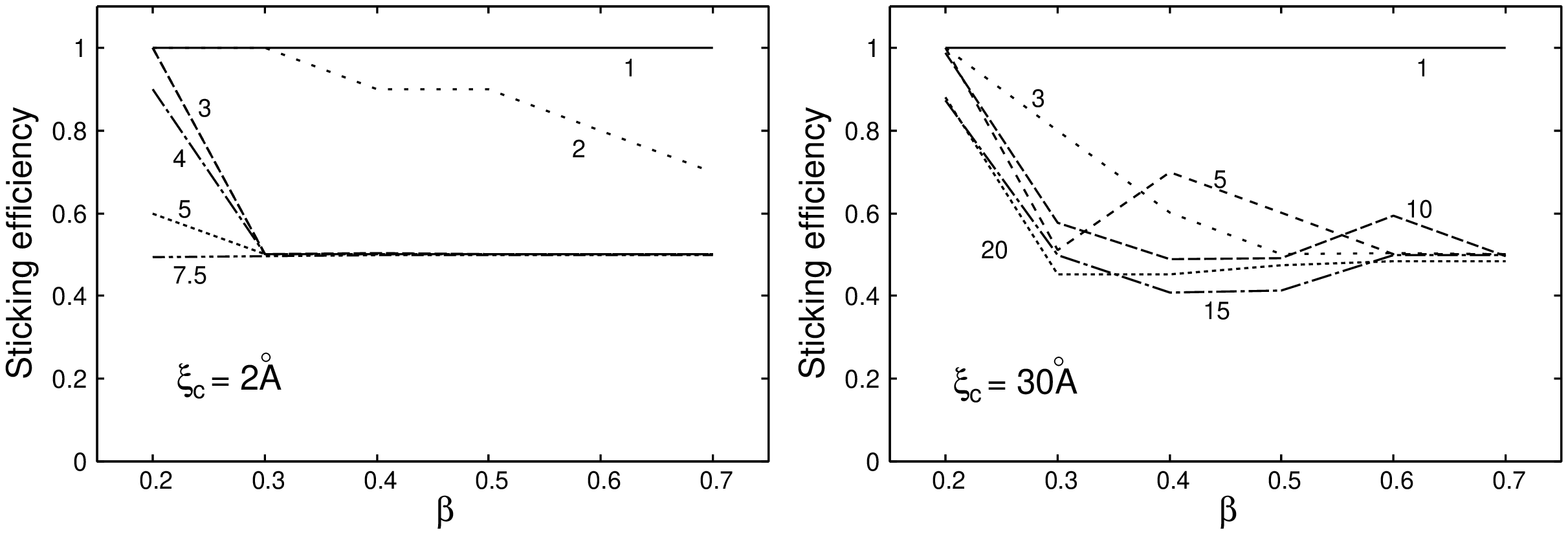}
\caption{Sticking efficiencies for compacted aggregates with various growth velocities {\bf with $\xi_{\rm c}=2$\AA (left) and $\xi_{\rm c}=30$\AA (right)}. Numbers denoted along each lines are $V_{\rm g}$ in ${\rm m\,s}^{-1}$ units. Data points are taken at an interval of $\Delta\beta =0.1$ and connected for ease of viewing.}
\label{fig:iv-beta}
\end{center}
\end{figure}

Compacted aggregates are produced through successive collisions with a collision velocity of $V_{\rm g}$ using non-sintered aggregates. The compacted aggregates drift to the central star and may enter a sintering zone. It is thus interesting to investigate the case in which $V_{\rm c}=V_{\rm g}$ for sintered aggregates. Figure~\ref{fig:iv-beta} shows the $\beta$ dependence of the sticking efficiency for aggregates grown at $V_{\rm g}$ colliding with $V_{\rm c}=V_{\rm g}$ {\bf for $\xi_{\rm c}=2$\,\AA (left) and 30\,\AA (right)}. If $V_{\rm c}=1\,{\rm m\,s}^{-1}$, the collisional outcome is perfect sticking, irrespective of $\beta$ {\bf in both cases. If $\xi_{\rm c}=2$\,\AA,} the fraction of bouncing increases as $\beta$ increases when $V_{\rm c}=2\,{\rm m\,s^{-1}}$. For $3\le V_{\rm c}\le 5\,{\rm m\,s^{-1}}$, only bouncing is observed except when $\beta=0.2$. At $\beta=0.2$, the fraction of bouncing increases as $V_{\rm g}$ increases. Bouncing at $\beta=0.2$ occurs when  $V_{\rm c}=7.5\,{\rm m\,s^{-1}}$. For $\xi_{\rm c}=30$\,\AA, the probability of sticking is slightly higher than $\xi_{\rm c}=2$\,\AA cases.

{\bf From these results, it is suggested that bouncing is one of the main collisional outcomes of compacted aggregates.} It is important to understand the conditions of bouncing in order to clarify the collisional evolution of icy grain aggregates. The conditions will be discussed in Section~5.4.

\begin{figure}[h]
\begin{center}
\includegraphics*[width=12cm,height=8cm]{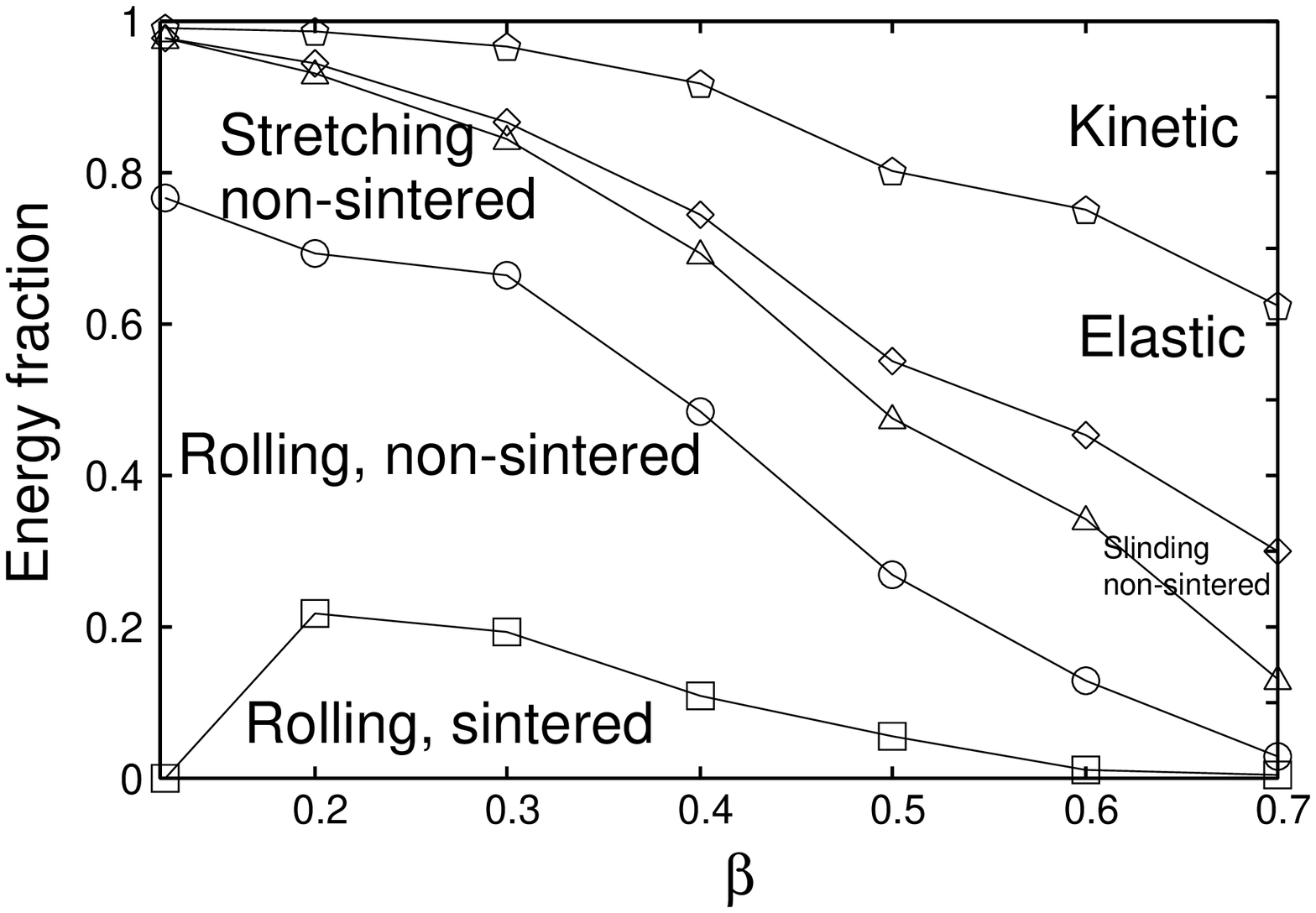}
\caption{Energy partition after collision of BCCA aggregate with a collision velocity of $V_{\rm c}=10\,{\rm m\,s}^{-1}$ as a function of $\beta$. Leftmost points are for non-sintered aggregates.}
\label{fig:energy}
\end{center}
\end{figure}

Figure~\ref{fig:energy} shows the partition of energy after the collision of BCCA aggregates with $V_{\rm c}=10\,{\rm m\,s}^{-1}$ as a function of $\beta$. {\bf The initial kinetic energy} dissipates through breaking of the sintered bonds, sliding and rolling friction at non-sintered contacts, and elastic vibration. Some of the kinetic energy remains after the collision. In Figure~\ref{fig:energy}, the leftmost points are non-sintered aggregates. Approximately 80\% of the initial kinetic energy dissipates through rolling friction, which causes the cause of smooth ductile deformation of the aggregates. The other 20\% is dissipated by breaking non-sintered necks through stretching. The  remaining kinetic energy is small, representing perfect sticking.

At $\beta=0.2$, 20\% of the kinetic energy dissipates through breaking sintered necks with rolling deformation. As $\beta$ increases, the amount of energy dissipation through the breaking by rolling decreases. Energy dissipation through rolling friction decreases accordingly. This is because the number of newly formed non-sintered contacts decreases as the number of broken sintered contacts  decreases. In turn, the elastic and remaining kinetic energies increase as $\beta$ increases. From Figure~\ref{fig:v10v30beta-s}, it can be seen that bouncing starts from $\beta=0.5$. At this value, 40\% of the initial kinetic energy does not dissipate but remains as kinetic and elastic energies. As a result, the efficiency of energy dissipation decreases and the collisional outcome becomes bouncing.

\begin{figure}[h]
\begin{center}
\includegraphics*[width=6cm,height=4cm]{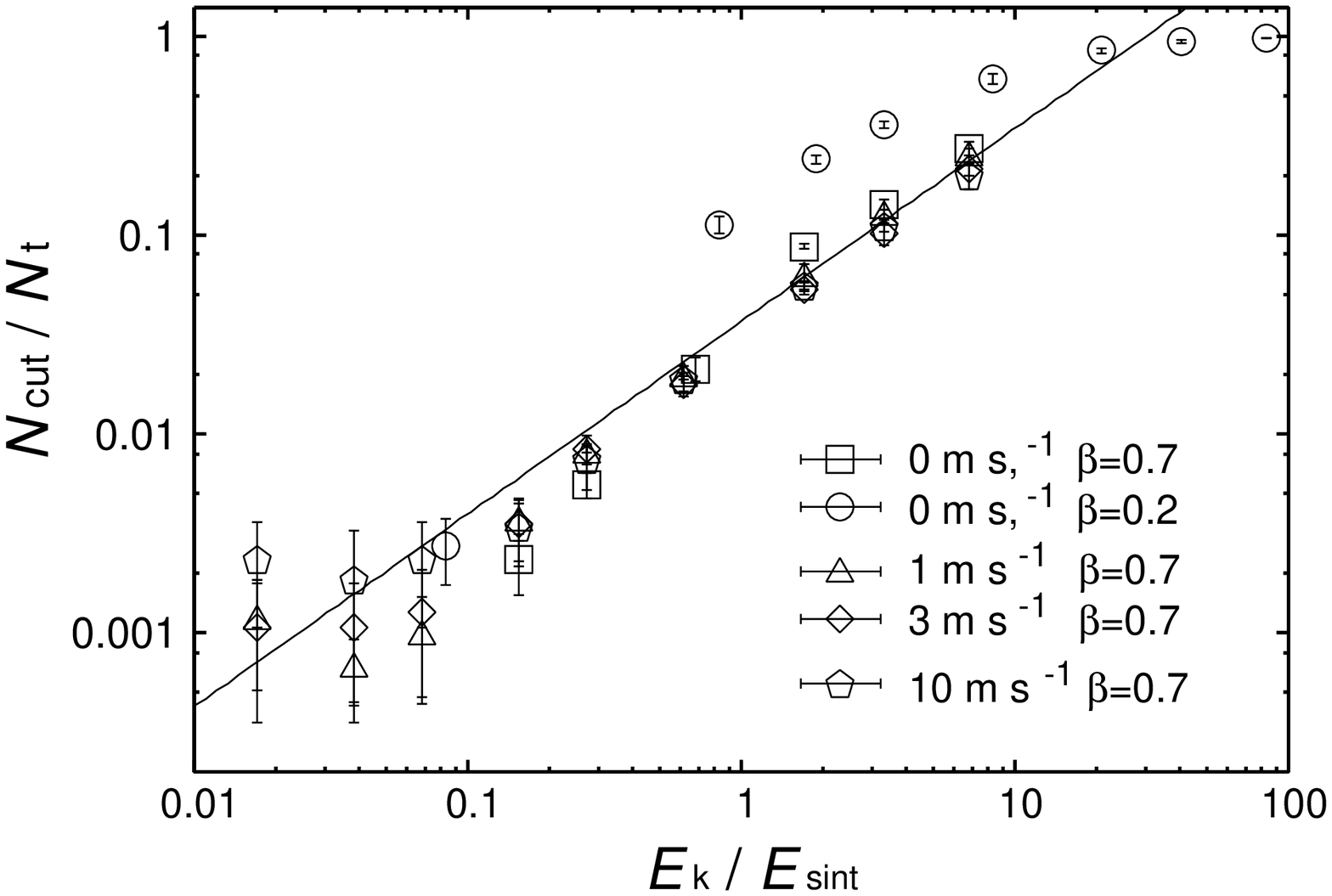}
\includegraphics*[width=6cm,height=4cm]{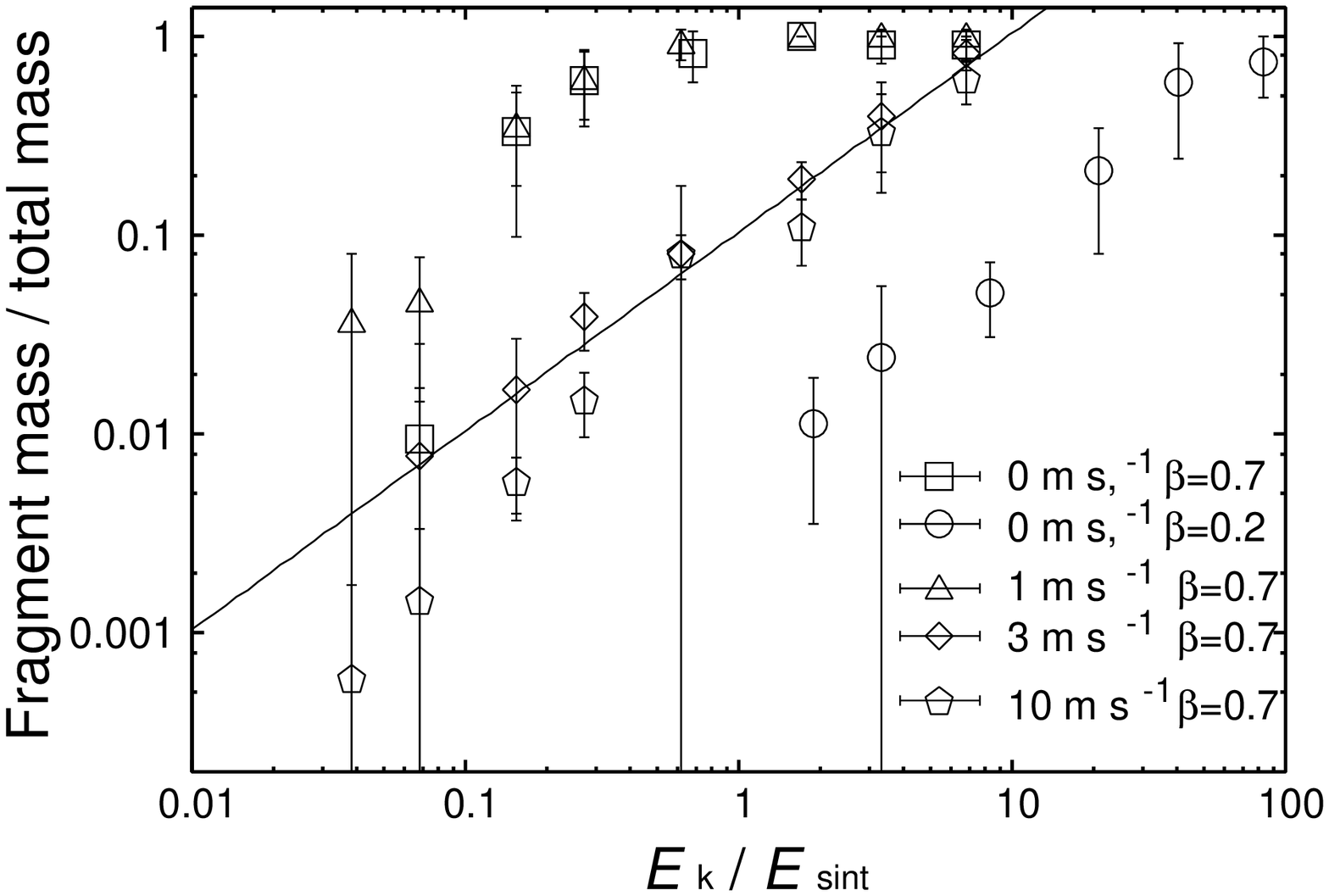}
\includegraphics*[width=6cm,height=4cm]{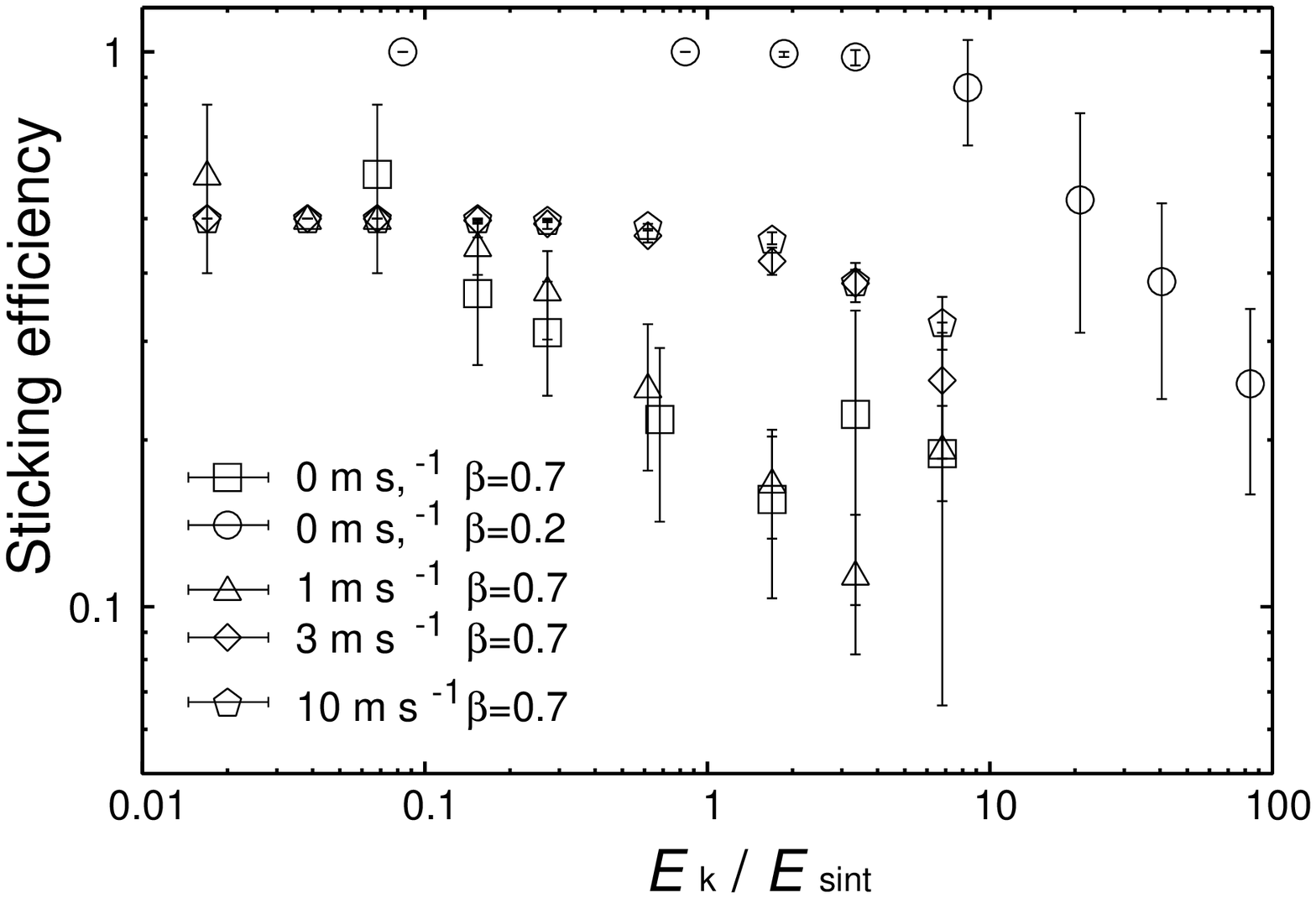}
\caption{Upper left: The number of broken contacts normalized by the total number of grains $N_{\rm cut}/N_{\rm t}$ as a function of the initial kinetic energy per grain normalized by the breaking energy for a sintered neck, $E_{\rm k}/E_{\rm sint}$.  Upper right: {\bf The fraction of mass in fragments compared to the total mass} as a function of $E_{\rm k}/E_{\rm sint}$. Bottom center: Sticking efficiency as a function of $E_{\rm k}/E_{\rm sint}$.  In every panel, two BCCA aggregates having $\beta=0.7$ (squares) and 0.2 (circles) and compacted aggregates with $V_{\rm g}=1$ (triangles), 3 (diamonds), and 10$\,{\rm m\,s}^{-1}$ (pentagons) are shown. {\bf Error bars are the standard deviations for five runs using five different aggregates (squares and circles) and using five different orientations (triangles, diamonds and pentagons).}}
\label{fig:cut}
\end{center}
\end{figure}

As shown above, the number of sintered necks broken through rolling motion is an important quantity in collision. Figure~\ref{fig:cut} left shows the number of sintered bonds broken in a collision normalized by the total number of grains, $N_{\rm cut}/N_{\rm t}$, as a function of the initial kinetic energy per unit grain, $E_{\rm k}=mV_{\rm c}^2/4$, normalized by the breaking energy of a sintered neck, $E_{\rm sint}=\pi\gamma (\beta R)^2$. Five types of aggregates are included: BCCA with $\beta=0.2$ and 0.7, and compacted aggregates with $V_{\rm g}=1$, 3, and 10\,m\,s$^{-1}$. Although all modes of motion are included (rolling, sliding, stretching) as a cause of breaking a sintered neck in Figure~\ref{fig:cut}, left, more than 90\% of the bonds were broken through rolling motion. It should be noted that data for all types of aggregates showed the same dependence except BCCA with $\beta=0.2$. Least-squares fitting except $\beta=0.2$ gave $3.7\times 10^{-2} (E_{\rm k}/E_{\rm sint})^{0.97}$, indicating that the number is determined by the ratio $E_{\rm k}/E_{\rm sint}$. The number of broken bonds thus depends only on  $\beta$, but not on the degree of compaction. 

On the contrary, {\bf the fraction of mass in fragments compared to the total mass} depends on the degree of compaction, as seen in Figure~\ref{fig:cut}, right. The fragment fraction was the largest for BCCA with $\beta=0.7$ and for compacted aggregates with $V_{\rm g}=1\,{\rm m\,s}^{-1}$. These two types of aggregates show the almost the same dependence. For high collision velocities, these aggregates are disrupted totally, as shown in Figure~\ref{fig:BCCA-s}~left. On the other hand, more compacted aggregates and BCCA with $\beta=0.2$ produced fewer fragments. The collisional outcomes of BCCA aggregates with $\beta=0.2$ are close to perfect sticking at collision velocities less than $32\,{\rm m\,s}^{-1}$, where $E_{\rm k}/E_{\rm sint}=8.3$, which is almost identical to the value for non-sintered aggregates, as seen before (Section~5.1). For compacted aggregates of $V_{\rm g}=3\,{\rm m\,s}^{-1}$ and $10\,{\rm m\,s}^{-1}$, the collisional outcome is bouncing. The {\bf mass} of fragments of these aggregates is between those of the other two types of aggregates. It can be seen that the fragment {\bf mass} is different for aggregates with the same $\beta$. Less-compacted aggregates produced more fragments even when the number of broken necks was the same (Figure~\ref{fig:cut}, {\bf upper right}). Correspondingly, the sticking efficiency (the mass of the largest aggregate) differs between the aggregate types (Figure~\ref{fig:cut},{\bf bottom}). This suggests that the fragments stick again to the main part of the aggregate during collision if the aggregate is compacted. Fragments are highly likely to collide with nearby fragment because the packing fraction is high. In contrast, fragments produced inside a porous aggregate have a chance to escape from the other fragments. Thus, grain arrangement is a critical parameter that determines the final mass of the largest aggregate. 

%Although both the number of broken necks and fraction of fragments are scaled with the initial kinetic energy, their dependences on the compaction degree are different. For $\beta=0.7$

\subsection{Conditions for bouncing}

The collisional outcomes can be categorized as sticking, bouncing, and fragmentation. It is helpful to clarify the conditions of collisional outcomes. 

In a non-sintered case, the condition given by \cite{Dominik} well describes the collisional outcomes. For example, maximum compression of an aggregate occurs when $E_{\rm roll}\simeq E_{\rm k}$, where $E_{\rm k}=mV^2/4$ is the kinetic energy of a grain and $E_{\rm roll}$ is the energy required to roll the grain 90$^\circ$ around an adjacent grain.  Catastrophic disruption occurs when $E_{\rm k}\simeq 10E_{\rm break}$, where $E_{\rm break}$ is the energy required to break a non-sintered contact. These considerations based on energetics were successful because a non-sintered aggregate smoothly deforms like liquid. This behavior comes from the fact that the force required for rolling motion is substantially smaller than those for other motions including sliding and stretching. As a result, smooth ductile deformation of a non-sintered aggregate is possible.  Energy thus plays a key role in the non-sintered case.

\begin{figure}[h]
\begin{center}
\includegraphics*[width=12cm,height=4cm]{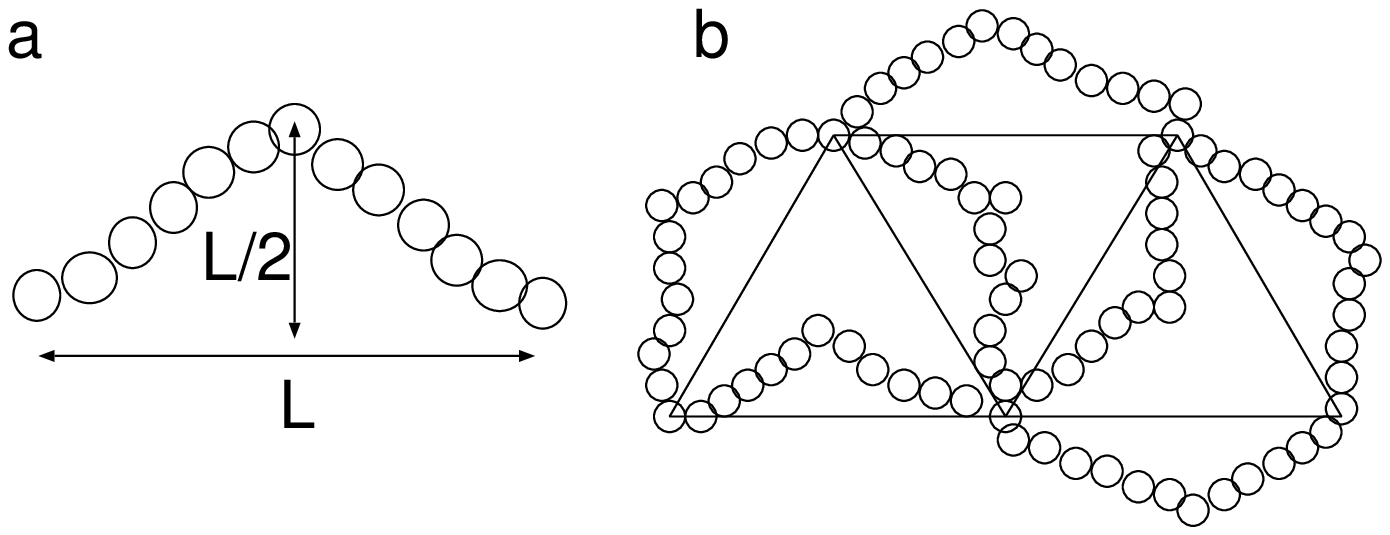}
\caption{(a) The basic structure of a grain aggregate. (b) Triangular chains of grains periodically connected to form triangle lattices.}
\label{fig:triangle}
\end{center}
\end{figure}

We showed that the number of broken bonds clearly depends on the ratio between the initial kinetic energy and the breaking energy of a sintered contact in Figure~\ref{fig:cut}.  A substantial fraction of sintered contacts are broken when $E_{\rm k}\simeq 10E_{\rm sint}$. Critical energies play a key role in fragmentation accordingly, as in the non-sintered case. However, the final mass of the largest aggregate depends on the packing fraction of an aggregate (Figure~\ref{fig:cut}~bottom). The arrangement of grains determines the  mass of the maximum aggregate. 

The situation is different for bouncing, because bouncing occurs if the force at a newly formed non-sintered contact is higher then the critical force for breaking a non-sintered contact. We investigated how forces are induced in an aggregate in order to clarify bouncing, adopting a model given by \cite{sirono00}, where an aggregate is represented by chains connected periodically. If we assume the shape of a chain is triangular with a base length of $L$ and a height $L/2$ (see Figure~\ref{fig:triangle}(a)), the number of grains in one chain, $N$, is given by
\begin{equation}
N=\left[\left({L\over 2R-1}\right)+\left(L\over 2R\right)\right]+1.
\label{eq:n}
\end{equation}
Suppose the chains are connected to form a regular triangular lattice (Figure~\ref{fig:triangle}(b)). It should be noted that the two end grains in one chain are shared with six triangular cells. $N-2$ in-between grains are sheared with the two neighboring  cells. In one cell, there are three end grains and $3\times (N-2)$ in-between grains. Thus, the number of grains contained in a unit cell is
\begin{equation}
N_{\rm cell}=3\times {1\over 6}+3\times {N-2\over 2}={3N-5\over 2}.
\label{eq:ncell}
\end{equation}

An end grain of a chain is shared with six chains. There are thus six contact points on an end grain. In one cell, there are three end grains having six contact points shared by six cells. There are three contact points contributed by end grains effectively in one cell. An in-between grain has two contact points. In one triangular cell, in-between grains contribute $2\times (N-2)/2\times 3=3(N-2)$ contact points. In total, there are $3N-3$ contact points for each cell. Thus, the average number of contact points on a grain is given by 
\begin{equation}
N_{\rm con}={6N-6\over 3N-5}.
\label{eq:ncon}
\end{equation}
The average number approaches to 2 as $N$ increases. 

%Figure\ref{fig:initn} shows the average number of contacts for compacted aggregates. We can obtain typical number of grains belonging to a chain in a compacted aggregate from  Fig.\ref{fig:initn} and $N_{\rm con}=(6N-6)/(3N-5)$.

\begin{figure}[h]
\begin{center}
\includegraphics*[width=7cm,height=5cm]{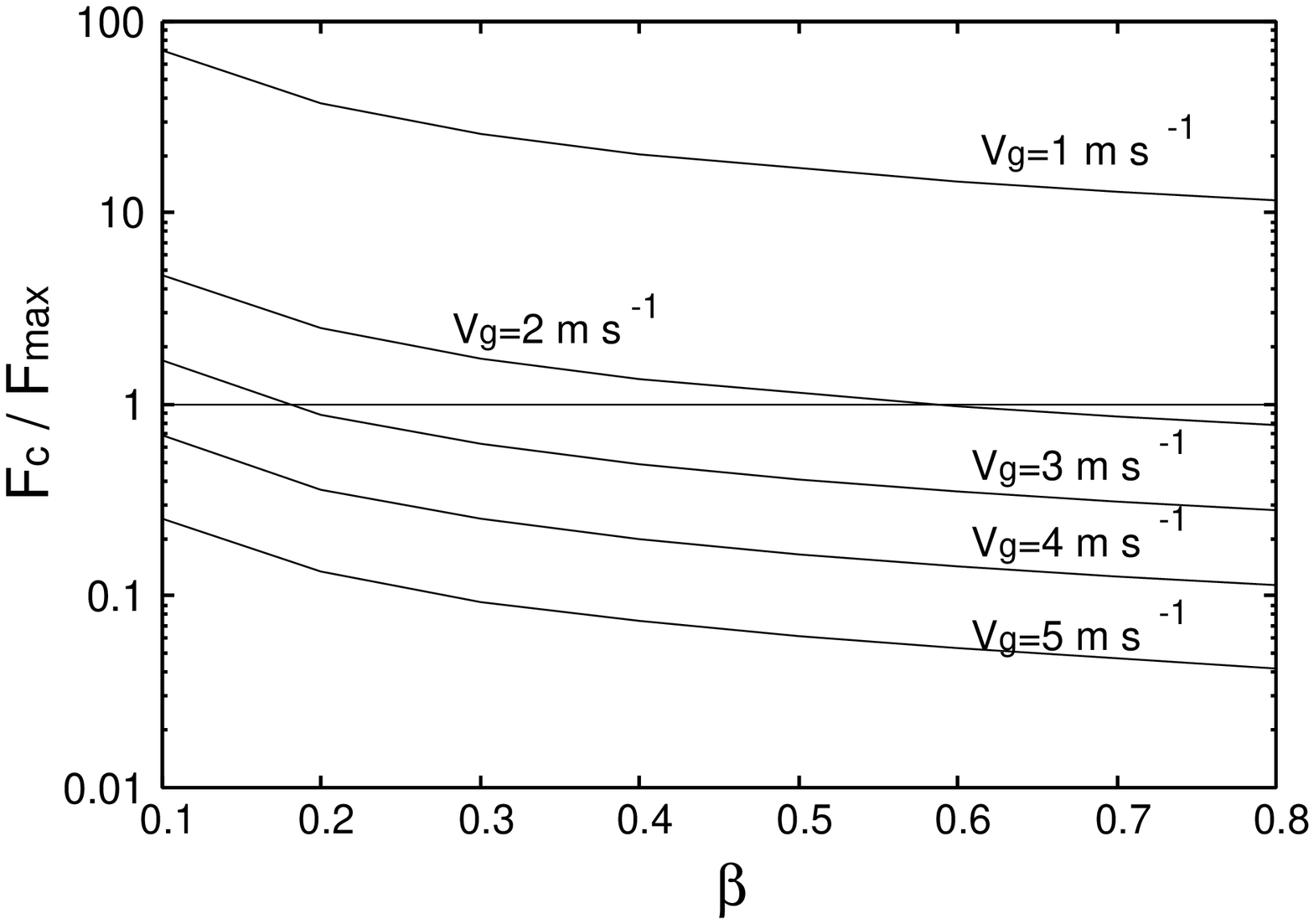}
\includegraphics*[width=7cm,height=5cm]{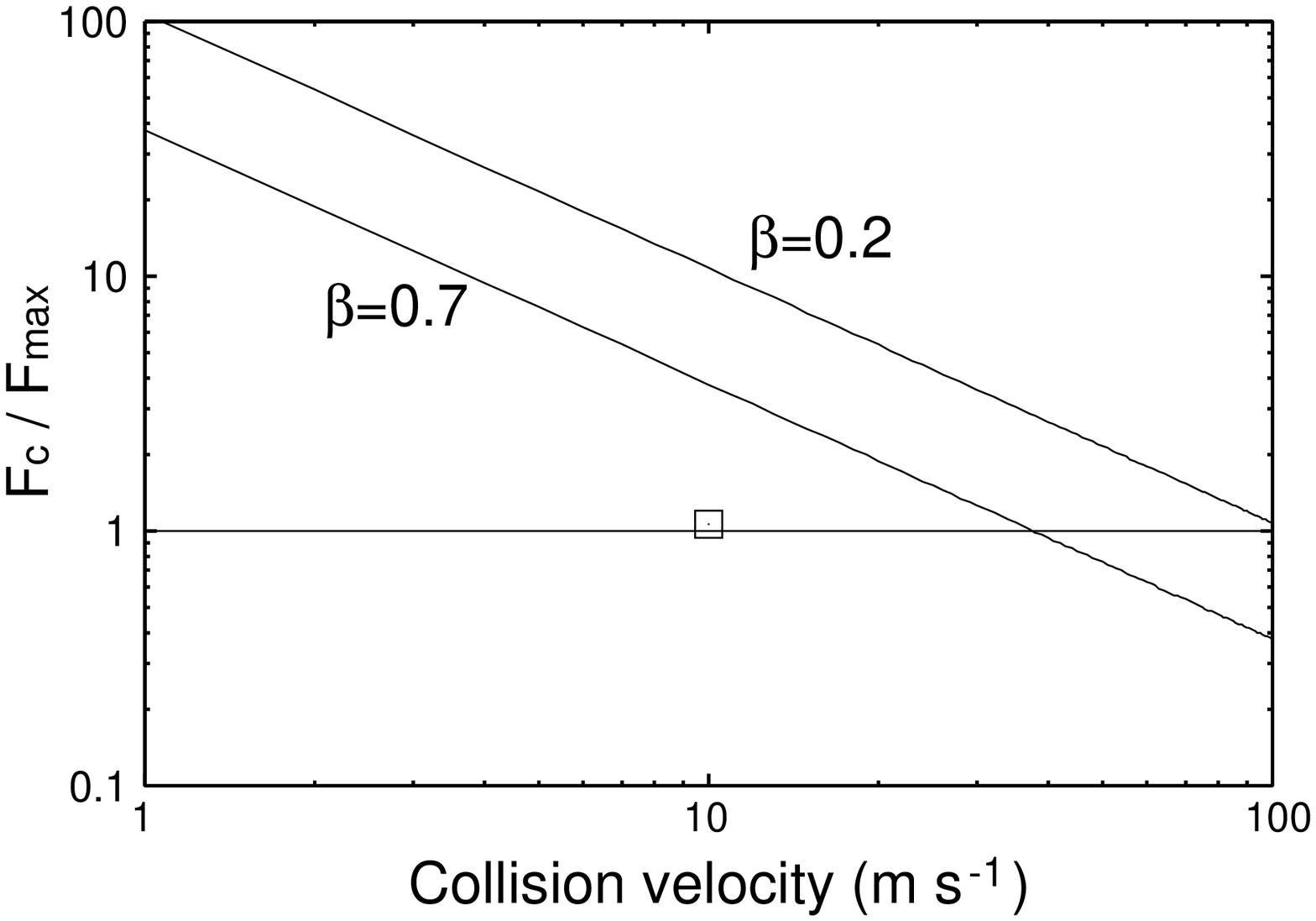}
\caption{Left: The ratio $F_{\rm c}/F_{\rm max}$ as a function of $\beta$ for different degree of compaction produced with $\xi_{\rm c}=2$\,\AA. From top to bottom, the growth velocity (=collision velocity) increases from 1\,m\,s$^{-1}$ to 5\,m\,s$^{-1}$. The horizontal line is $F_{\rm c}/F_{\rm max}=1$, below which bouncing is expected. Right: The ratio $F_{\rm c}/F_{\rm max}$ as a function of collision velocity for BCCA aggregates with $\beta=0.2$ and 0.7. The square is the ratio estimated taking new contact formation into account.}
\label{fig:ratio}
\end{center}
\end{figure}

When aggregates collide, forces are induced at the contacts. The maximum force induced in an aggregate is determined by the balance of kinetic energy and elastic energy stored in the chains. If the total number of chains in an aggregate is $N_{\rm chain}$, the balance can be written by $N_{\rm chain}k_{\rm chain}u_{\rm chain}^2/2=N_{\rm g}V_{\rm c}^2/4$, where $N_{\rm g}$ is the number of grains in the aggregate, and $u_{\rm chain}$ is the maximum shrinkage of the chain. The effective spring constant of the triangular chain $k_{\rm chain}$ is given  by \citep{sirono00} as
\begin{equation}
k_{\rm chain}={12(N-1)k_{\rm sr}\over (N-3)(N+1)}\left({L\over 2R-1}\right)^{-2},
\label{eq:kg}
\end{equation}
where the term $k_{\rm sr}=\pi(\beta R)^4E/4d$ appeared in Equation~(\ref{eq:M}). Because $N_{\rm g}/N_{\rm chain}\simeq N$, the balance gives a maximum force $F_{\rm max}$ as
\begin{equation}
F_{\rm max}=V_{\rm c}\sqrt{k_{\rm chain}mN/2}.
\label{eq:fmax}
\end{equation}
When two aggregates collide, the maximum compressive force given by Equation~(\ref{eq:fmax}) is induced inside an aggregate. At the same time, non-sintered new contacts are formed during the collision. An aggregate bounces if the tensile force at the new contact is larger than the critical force for breaking a non-sintered contact. Because the maximum compressive and tensile forces are comparable, if $F_{\rm max}$ is larger than the critical force for breaking, $F_{\rm c}=3\pi \gamma R^*$, bouncing occurs. This condition can be written as
\begin{equation}
{F_{\rm c}\over F_{\rm max}}<1.
\label{eq:ratio}
\end{equation}
Figure~\ref{fig:ratio} shows the ratio given by Equation~(\ref{eq:ratio}) corresponding to compact aggregates produced with  $\xi_{\rm c}=2$\,\AA (left) and BCCA aggregates (right). To calculate $F_{\rm max}$, we have to adopt $N$ into Equation~(\ref{eq:fmax}). For BCCA aggregates, we determined $N$ from $L=100R$ from the gyration radius shown in Figure~\ref{fig:BCCA-s}~right. For compacted aggregates, $N$ was determined from Equation~(\ref{eq:ncon}) and the average number of contacts shown in Figure~\ref{fig:initn} {\bf ($\xi_{\rm c}=2$\,\AA)}. In Figure~\ref{fig:ratio} left, the ratio was calculated for collisions of compacted aggregates with $V_{\rm c}=V_{\rm g}$.  

From Figure~\ref{fig:ratio} left, the model predicts that aggregates compacted with $V_{\rm g}=2\,{\rm m\,s^{-1}}$ colliding at the same velocity bounce if $\beta$ is larger than 0.4. Aggregates of $V_{\rm g}=1\,{\rm m\,s^{-1}}$ stick, irrespective of $\beta$, and aggregates of $V_{\rm g}=3\,{\rm m\,s^{-1}}$ bounce if $\beta\ge 0.2$. These results well explain the collisional outcomes shown in Figure~\ref{fig:iv-beta}{\bf, left ($\xi_{\rm c}=2$\,\AA)}.

For BCCA aggregates (Figure~\ref{fig:ratio}, right), the model predicts sticking for a wide range of collision velocities in contrast to the results shown in Figure~\ref{fig:BCCA-s}~left, where $\beta=0.7$ aggregates bounce at $V_{\rm c}=10\,{\rm m\,s}^{-1}$. It should be noted that there are no chain loops inside a BCCA aggregate because of its formation mechanism i.e., the motion of the colliding aggregate stops when a new contact is formed. 

When {\bf sintered} BCCA aggregates collide, they elastically deform substantially (Figure~\ref{fig:coltime}). A chain of grains bends and new contacts are formed around the impact point. The length of the chains decreases, and larger forces can appear as the elasticity of the aggregate increases. We identified loops consisting of grain chains and checked the the number of grains belonging to the loops. The average number of the grains was 44. We therefore adopted $N=44/2=22$ in Equation~(\ref{eq:fmax}). The square in Figure~\ref{fig:ratio}, right is the ratio  based on the calculated number of grains belonging to the loops. The ratio is almost unity in this case, well explaining the results obtained by numerical simulation (sticking efficiency of 0.6, corresponding to two bounces and three stickings in Figure~\ref{fig:BCCA-s}~left). 

As seen above, the threshold between bouncing and sticking is determined by forces induced in an aggregate and the sticking force needed to separate grains. This is in contrast to the collisions between non-sintered aggregates, in which the collisional outcomes are well classified based on energies associated with the  motions of the grains.

\subsection{Uncertainty in $\xi_{\rm c}$}
{\bf A critical displacement for irreversible rolling motion of $\xi_{\rm c}=2$\AA  was estimated by \cite{Dominik}}. On the other hand, \cite{Heim} obtained  $\xi_{\rm c}=32$\AA\, experimentally. Figure~\ref{fig:iv-beta} displays the effect of $\xi_{\rm c}$ on the collisional outcomes of compacted aggregates. As seen in Figure~\ref{fig:initn}, the number of contact points decreases as  $\xi_{\rm c}$ increases. The number of contacts at $V_{\rm g}=3\,{\rm m\,s}^{-1}$ with $\xi_{\rm c}=2$\AA\, is almost the same as that at $V_{\rm g}=10\,{\rm m\,s}^{-1}$ with $\xi_{\rm c}=30$\AA. The results of $V_{\rm g}=3\,{\rm m\,s}^{-1}$ with $\xi_{\rm c}=2$\AA\, in Figure~\ref{fig:iv-beta}~{\bf left} and $V_{\rm g}=10\,{\rm m\,s}^{-1}$ with $\xi_{\rm c}=30$\AA\, in Figure~\ref{fig:iv-beta}~{\bf right}, well coincides. This comparison shows that the number of contacts is a critical parameter, which in turn depends on $\xi_{\rm c}$.

On the other hand, Figure~\ref{fig:v10v30beta-s}, right, shows the sticking efficiencies of BCCA aggregates with $\xi_{\rm c}=30$\AA. Comparing Figure~\ref{fig:v10v30beta-s} {\bf left and right}, no significant change is observed. This is because the number of contacts in both aggregates is the same. Although new contacts that form during a collision are non-sintered ones and their rolling motion depends on $\xi_{\rm c}$, BCCA aggregates at high collisional velocities break up totally and the dependence on $\xi_{\rm c}$ is not observed.

\section{Discussion}

\subsection{Evolution of icy dust aggregates}
In the non-sintering zone of a protoplanetary nebula, an aggregate efficiently grows through collisional sticking, and the collision velocity increases as the aggregate size increases. At the same time, the aggregates are compacted. The compacted aggregates drift to the central star owing to gas drag. If an aggregate that has grown in a non-sintered zone drifts into a sintering zone, sintering occurs inside the aggregate and its mechanical properties change.  Figure~\ref{fig:iv-beta} suggests that compacted aggregates that grow with a velocity of $V_{\rm g}=3\,{\rm m\,s^{-1}}$ cannot grow further because of bouncing {\bf if $\beta$ is high enough}. 

On the other hand, aggregates that form inside a sintering zone grow being less compacted than those in a non-sintered zone (Figure~\ref{fig:BCCA-s}~right). Figure~\ref{fig:BCCA-s}~left shows that porous sintered aggregates break at a collision velocity of $20\,{\rm m\,s^{-1}}$. In this case, the size of the aggregate is limited by fragmentation, not by bouncing, as discussed for compacted aggregates. 

These two effects induce heterogeneity in the growth of icy dust aggregates in a protoplanetary nebula. In a non-sintered zone, perfect sticking of aggregates does not produce fragments in contrast to the sintered zone. This contrast can explain the symmetric pattern observed at HL Tau \citep{ALMA}. \cite{okuzumi} reproduced the observational pattern assuming substantial fragment production at collisional velocities higher than 20\,m\,s$^{-1}$. This velocity corresponds to the critical velocity for the growth of sintered BCCA aggregates. The effect of bouncing collisions and associated fragment production (Figure\ref{fig:v10v30beta-f}) should be included in future studies. 

\subsection{Limitations of the simulations}
The simulations conducted in this study were 2-D as a first step. Of course, 3-D simulations are also required. However, important aspects of sintered aggregate collisions are common in 2-D and 3-D.

Figure~\ref{fig:ratio} nicely explains the threshold between sticking and bouncing after a collision. This figure is based on a comparison between the critical pull-off force, $F_{\rm c}=3\pi \gamma R^*$, and the maximum force $F_{\rm max}$ (Equation~(\ref{eq:fmax})) attained in a collision. The pull-off force is common in 2-D and 3-D. How about the maximum force? The discussion in Section~5.4 is based on a comparison between the initial kinetic energy and the total elastic energy stored in an aggregate, which is independent of dimensionality. Thus, the conditions obtained in this study would be applicable also to 3-D cases. 

Moreover, we found that the number of broken sintered necks only depended on the initial kinetic energy (Figure\ref{fig:cut}, left). This point is also independent of dimensionality. Thus, the critical velocity at which fragmentation starts should be the same in both 2-D and 3-D cases. However, the {\bf mass} of fragments might depend on the dimension because the mass depends on the arrangement of grains  in the aggregate {\bf (Figure\ref{fig:cut}, right)}. 

Another important point is the number of grains in an aggregate. In this study, aggregates composed of 1024 grains were used. This number is quite small for simulating actual aggregates. However, the discussion in Section~5.4 suggests that the collisional outcome, especially sticking or bouncing, is determined by the length of the chains composing the aggregates. In other words, a critical parameter is the packing fraction of an aggregate. The geometrical meaning of the packing fraction is quite different between 2-D and 3-D. The probability of  a fragment escaping would be higher in 3-D than in 2-D. Thus, it is possible that the {\bf mass} of fragments shown in Figure~\ref{fig:cut}, upper right, would be higher in 3-D. This point should be addressed by 3-D simulation.

\subsection{Roughness of a broken neck}
As shown in Figure~\ref{fig:cut}, sintered necks are broken during collisions. The shape of a broken neck is not smooth; instead, it has edges. The collisional outcomes of sintered aggregates highly depend on the stickiness of fragments. A fragment is produced through breaking necks by rolling motion and the fragment collides with other fragments. If the fragment sticks, mass loss due to fragmentation is small. If the fragment bounces, the mass loss is large.

One criterion of bouncing is $F_{\rm c}/F_{\rm max}< 1$, where $F_{\rm c}$ is the pull-off force, given by $3\pi \gamma R^*$. If an aggregate is highly porous, the shape of a fragment is a chain. When a fragment chain collides, it is highly probable that the end of the chain is the first to come into contact with the other chain. Because the end of a chain is produced by breaking a neck, the contacting surface is likely rough. 

Although this effect is not included in this study,  we can discuss a consequence of this effect. Suppose the rough surface has a surface curvature of $0.1R$. The lines shown in Figure~\ref{fig:ratio}, right, would shift downward by a factor of 10. The region of bouncing is significantly reduced, and the collisional outcome tends to bouncing (or fragmentation). 

Another important effect not included in this study is the motion of the grain during rolling. After a neck is broken through rolling, a grain rolls around the axis at the edge of the neck (see Figure\ref{fig:tilt}). During rolling, the center of the rolling grain shifts upward because the distance between the axis and the center is larger than that between the contact surface and the center. As a result, the distance between two grain centers increases. Clearly this effect inhibits reconnection of the grains. This effect should be included in the future simulation.

\section{Conclusion}
We conducted numerical simulations of icy dust aggregate collisions, including the effects of sintering. Sintering increases the mechanical strength of the neck connecting adjacent grains and decreases the volume of a grain. We modeled the mechanical responses of the neck and included them in the simulation code. For very porous dust aggregates, sintering reduces the critical velocity for growth from 50 \,m\,s$^{-1}$ in the non-sintered case to 20\,m\,s$^{-1}$ in the sintered case. This is because a limited fraction of necks is broken during rolling of  grains in a sintered aggregate. This is not the case for non-sintered aggregates, where a substantial fraction of grains smoothly rolls around adjacent grains without breaking. For compacted aggregates, the collisional outcome is bouncing. The amount of fragments increases as the initial kinetic energy increases. The number of necks broken during a collision does not depend on the porosity of an aggregate. From the simulation results obtained in this study, it is suggested that sintering inhibits the growth of icy aggregates in their collisional evolution.

\acknowledgments

The authors greatly thank to the critical and constructive comments by an anonymous reviewer. This work was supported by JSPS KAKENHI Grant Number 17K05631.

\clearpage

%% the file, uncomment the \plotone, \plottwo, and
%% If you need a layout that cannot be achieved with \plotone or
%% \plottwo, you can invoke the graphicx package directly with the
%% \includegraphics command or use \plotfiddle. For more information,
%% please see the tutorial on "Using Electronic Art with AASTeX" in the
%% documentation section at the AASTeX Web site,
%% http://www.journals.uchicago.edu/AAS/AASTeX.
%%
%% The examples below also include sample markup for submission of
%% supplemental electronic materials. As always, be sure to check
%% the instructions to authors for the journal you are submitting to
%% for specific submissions guidelines as they vary from

%% This example uses \plotone to include an EPS file scaled to
%% 80% of its natural size with \epsscale. Its caption
%% has been written to indicate that additional figure parts will be
%% available in the electronic journal.

\clearpage

%% Here we use \plottwo to present two versions of the same figure,
%% one in black and white for print the other in RGB color
%% for online presentation. Note that the caption indicates
%% that a color version of the figure will be available online.
%%

%% This figure uses \includegraphics to scale and rotate the still frame

%% If you are not including electonic art with your submission, you may
%% mark up your captions using the \figcaption command. See the
%% No more than seven \figcaption commands are allowed per page,
%% so if you have more than seven captions, insert a \clearpage

%% Tables should be submitted one per page, so put a \clearpage before

%% Two options are available to the author for producing tables:  the
%% deluxetable environment provided by the AASTeX package or the LaTeX
%% table environment.  Use of deluxetable is preferred.
%%

%% Three table samples follow, two marked up in the deluxetable environment,
%% one marked up as a LaTeX table.

%% In this first example, note that the \tabletypesize{}
%% command has been used to reduce the font size of the table.
%% We also use the \rotate command to rotate the table to
%% landscape orientation since it is very wide even at the
%% Note also that the \label command needs to be placed

%% This table also includes a table comment indicating that the full
%% version will be available in machine-readable format in the electronic

\clearpage
%% Text for table notes should follow after the \enddata but before
%% the \end{deluxetable}. Make sure there is at least one \tablenotemark
%% in the table for each \tablenotetext.

%% If you use the table environment, please indicate horizontal rules using
%% Do not put multiple tabular environments within a single table.
%% The optional \label should appear inside the \caption command.

\clearpage

%% Any table notes must follow the \end{tabular} command.

%% If the table is more than one page long, the width of the table can vary
%% from page to page when the default \tablewidth is used, as below.  The
%% individual table widths for each page will be written to the log file; a
%% maximumtablewidth for the table can be computed from these values.
%% The \tablewidth argument can then be reset and the file reprocessed, so
%% that the table is of uniform width throughout. Try getting the widths
%% from the log file and changing the \tablewidth parameter to see how
%% adjusting this value affects table formatting.

%% The \dataset{} macro has also been applied to a few of the objects to
%% show how many observations can be tagged in a table.

\clearpage

%% Tables may also be prepared as separate files. See the accompanying
%% sample file table.tex for an example of an external table file.
%% To include an external file in your main document, use the \input
%% command. Uncomment the line below to include table.tex in this
%% sample file. (Note that you will need to comment out the \documentclass,
%% \begin{document}, and \end{document} commands from table.tex if you want
%% to include it in this document.)


%% The following command ends your manuscript. LaTeX will ignore any text


\begin{thebibliography}{}
\bibitem[ALMA Partnership et al.(2015)]{ALMA} ALMA Partnership, Brogan, C. L., P\'erez, L. M., et al. 2015, \apj, 808, L3
\bibitem[Blackford(2007)]{sintrev} Blackford, J. R. 2007,  J. Phys. D:Appl. Phys, 40, R355
\bibitem[Chokshi et al.(1993)]{Chokshi} Chokshi, A., Tielens, A. G. G. M., \& Hollenbach, D. 1993, \apj, 407, 806
\bibitem[Dominik \& Tielens(1995)]{Dominik95} Dominik, C., \& Tielens, A. G. G. M. 1995, Phil. Mag. A, 72, 783 
\bibitem[Dominik \& Tielens(1996)]{Dominik96} Dominik, C., \& Tielens, A. G. G. M. 1996, Phil. Mag. A, 73, 1279 
\bibitem[Dominik \& Tielens(1997)]{Dominik} Dominik, C., \& Tielens, A. G. G. M. 1997, \apj, 480, 647
\bibitem[Krijt et al.(2013)]{Krijt13} Krijt, S., G\"uttler, C., Hei\ss elmann, D. Dominik, C., \& Tielens, A. G. G. M. 2013, J. Phys. D: Appl. Phys., 46, 435303
\bibitem[Krijt et al.(2014)]{Krijt14} Krijt, S.,  D. Dominik, C., \& Tielens, A. G. G. M. 2013, J. Phys. D: Appl. Phys., 47, 175302
\bibitem[Kuroiwa \& Sirono(2011)]{kuroiwa} Kuroiwa, T., \& Sirono, S. 2011, \apj, 739, 18 
\bibitem[Gibb et al.(2004)]{Gibb} Gibb, E. L., Whittet, D. C. B., Boogert, A. C. A., \& Tielens, A. G. G. M. 2004, \apj, 151, 35
\bibitem[Heim et al.(1999)]{Heim} Heim, L.-O., Blum, J., Preuss, M., \&
		Butt, H.-J. 1999, Phys. Rev. Lett., 83, 3328
\bibitem[Johnson (1987)]{Contact} Johnson, K. L. 1987, Contact mechanics, (Cambridge, Cambridge Univ. Press)
\bibitem[Kataoka et al.(2013)]{kataoka} Kataoka, A., Tanaka, H., Okuzumi, S., \& Wada, K. 2013, \aap, 557, L4
\bibitem[Kimura et al.(2015)]{Kimura} Kimura, H., Wada, K., Sensyu, H., \& Kobayashi, H. 2015, \apj, 812, 67
\bibitem[Maeno \& Ebinuma(1983)]{Maeno} Maeno, N., \& Ebinuma, T. 1983, J. Phys. Chem. 87, 4103
\bibitem[Meakin(1991)]{Meakin} Meakin, P. 1991, Rev. Geophys., 29, 317
\bibitem[Okuzumi et al.(2012)]{Okuzumi12} Okuzumi, S., Tanaka, H., Kobayashi,  H., \& Wada, K. 2012, \apj, 752, 106
\bibitem[Okuzumi et al.(2016)]{okuzumi} Okuzumi, S., Momose, M., Sirono, S., Kobayashi, H.,\& Tanaka, H. 2016, \apj, 821, 82
\bibitem[Ormel \& Cuzzi(2007)]{Ormel} Ormel, C. W., \& Cuzzi, J. N. 2007, \aap, 466, 413
%\bibitem[Petrovic(2003)]{Petrovic} Petrovic, J. J. 2003, J. Mat. Sci., 38, 1
\bibitem[Poppe et al.(2000)]{Poppe} Poppe, T., Blum, J., \& Henning, T. 2000, \apj, 533, 454
\bibitem[Sirono(1999)]{Sirono} Sirono, S. 1999, \aap, 347, 720
\bibitem[Sirono \& Greenberg(2000)]{sirono00} Sirono, S., \& Greenberg, J. M. 2000, \icarus, 145, 230
\bibitem[Sirono(2011)]{sirono11} Sirono, S. 2011, \apj, 735, 131 
\bibitem[Wada et al.(2007)]{Wada07} Wada, K., Tanaka, H., Suyama, T.,
                Kimura, H. \& Yamamoto, T. 2007, \apj, 661, 320
\bibitem[Wada et al.(2009)]{Wada} Wada, K., Tanaka, H., Suyama, T.,
                Kimura, H. \& Yamamoto, T. 2009, \apj, 702, 1490
\bibitem[Weidenschilling (1977)]{wei77} Weidenschlling, S. J. 1977, \apss, 51, 153
%\bibitem[Li \& Wang(1999)]{Li} Li, W., \& Wang, T. 1999. Phys. Rev. B, 59, 3993
%\bibitem[Yang et al.(2005)]{Yang} Yang, W., Araki, H., Tang, C., Thaveethavorn, S., Suzuki, H., \& Noda T. 2005, Adv. Mater., 17, 1519
\end{thebibliography}
\end{document}